\documentclass{aa}  
\usepackage{scrextend}
\usepackage{hyperref}
\usepackage{pdflscape}
\usepackage{xcolor}
\usepackage{rotating}
\usepackage{graphicx}
\usepackage{soul}
\usepackage{amsmath}

\usepackage{txfonts}
\usepackage{siunitx}
\usepackage{enumitem}
\usepackage{multirow}
\usepackage{lscape}
\usepackage{longtable}
\usepackage{tikz}
\usepackage{subfigure}

\usepackage{booktabs}

\usepackage{cleveref}
\usepackage[footnotesize]{caption}
\usepackage[flushleft]{threeparttable}

\newcommand{\Msun}{$M_{\odot}$}

\newcommand{\qout}{$q_{\mathrm{out}}$}
\newcommand{\eout}{$e_{\mathrm{out}}$}
\newcommand{\ain}{$a_{\mathrm{in}}$}
\newcommand{\aout}{$\tilde{a}_{\text{out}}$}

\newcolumntype{R}{@{\extracolsep{3cm}}r@{\extracolsep{0pt}}}
\usepackage{graphicx,caption,subcaption}
\usepackage{txfonts}
\begin{document}

   \title{Southern Massive Stars at High Angular Resolution (SMaSH+): Properties of hierarchical massive triples\thanks{Based on observations collected at the European Organisation for Astronomical Research in the Southern Hemisphere under ESO programme 189.C-0644.}}

   \author{E. Bordier\inst{\ref{kul},\ref{koeln}}, H. Sana\inst{\ref{kul},\ref{lgi}}, A. J. Frost\inst{\ref{eso}}, A.-S. Libert\inst{\ref{Unamur}}, J. Vrancken\inst{\ref{kul}}, L. Mahy\inst{\ref{rob}},  S. Toonen\inst{\ref{api},\ref{grappa}}, F. Tramper\inst{\ref{cab}}, A. de Koter\inst{\ref{kul},\ref{api}}, S. Lacour\inst{\ref{paris}}, J.-B. Le Bouquin\inst{\ref{grenoble}}, W.-J. de Wit\inst{\ref{eso}}}

\institute{  Institute of Astronomy, KU Leuven, Celestijnenlaan 200D, B-3001 Leuven, Belgium\label{kul} \email{bordier@ph1.uni-koeln.de} \and  
I. Physikalisches Institut, Universität zu Köln, Zülpicher Str. 77, Cologne, 50937, Germany \label{koeln} \and 
Leuven Gravity Institute, KU Leuven, Celestijnenlaan 200D box 2415, 3001 Leuven, Belgium\label{lgi} \and
European Southern Observatory, Alonso de C\'ordova 3107, Casilla 19, Santiago, Chile\label{eso} \and
naXys Research Institute, Department of Mathematics, University of Namur, 61 Rue de Bruxelles, 5000 Namur, Belgium \label{Unamur} \and  
Royal Observatory of Belgium, Avenue Circulaire/Ringlaan 3, 1180 Brussels, Belgium\label{rob} \and 
Anton Pannekoek Institute for Astronomy, University of Amsterdam, 1090 GE Amsterdam, The Netherlands\label{api} \and
GRAPPA, University of Amsterdam, 1090 GE Amsterdam, The Netherlands \label{grappa} \and
Centro de Astrobiolog{\'i}a (CAB), CSIC-INTA, Carretera de Ajalvir km 4, E-28850 Torrej{\'o}n de Ardoz, Madrid, Spain\label{cab} \and 
LESIA, Observatoire de Paris, PSL, CNRS, Sorbonne Université, Université de Paris, 5 place Janssen, 92195 Meudon, France\label{paris} \and  
Univ. Grenoble Alpes, CNRS, IPAG, 38000 Grenoble, France\label{grenoble} }

   \date{Received Month xx, xxxx; accepted Month xx, xxxx}

    \titlerunning{Properties of hierarchical massive triples in SMaSH+}
   \authorrunning{E.~Bordier et al.}

 
  \abstract
   {While massive stars are frequently found in triple architectures, the lack of observed parameter distributions has long remained a bottleneck for statistical models of their evolution.}
   {We bridge this gap by compiling the first representative set of physical and orbital distributions for main-sequence hierarchical massive triples.}
   {We present a homogeneous analysis of 26 O-type hierarchical triples identified in the SMaSH+ survey by combining spectroscopic data for inner binaries with interferometric and aperture masking detections of tertiary companions within $\sim 200$~au. We derive the distributions of masses, mass ratios, and separations, and investigate their joint probability density functions. We assess the dynamical stability of these systems. We estimate the relative importance of secular processes by comparing the von Zeipel–Kozai–Lidov (ZKL) timescale to the general relativistic precession timescale for several systems with well-constrained orbital solutions. Finally, we evaluate the observational completeness and find no substantial detection bias for separations $\lesssim 100$~au and mass ratios \qout>0.1.}
   {The sample is dominated by strongly hierarchical configurations, consisting primarily of tight inner spectroscopic binaries (\ain$<1$~au) and wider tertiaries ($\frac{\tilde{a}_{\mathrm{out}}}{a_{\mathrm{in}}}>70$ for most systems). We find no significant correlation between tertiary mass and either inner-binary mass or outer separation, indicating a broad diversity of system architectures. Ten systems (out of 26) host relatively massive tertiaries (\qout$>$0.5), especially at closer outer separations (\aout$\lesssim$30~au). For two to four systems out of five with well-constrained orbits, general relativistic precession dominates over ZKL oscillations in their current configuration, although ZKL-driven migration may have contributed to the formation of the observed short-period inner binaries. These results provide the first observationally grounded distributions of key parameters for massive hierarchical triples and offer important constraints for population synthesis and evolutionary models, particularly regarding the role of tertiary companions in shaping binary evolution.}
   {}


\maketitle

\section{Introduction}
High multiplicity is a well-established characteristic of massive stars and a central factor in understanding their formation and evolution. Over the last decade, observational studies have shown that the majority of massive stars will interact with at least one companion during their lifetimes \citep{DeWit+2005,Sana+2012,Sana+2013,deMink+2014,Banyard+2022}. 
With the advancement of high angular resolution instruments, the observational paradigm now indicates that O- and B-type stars frequently reside in higher-order configurations, where triple systems represent a significant fraction of the population \citep{Sana+2014,Moe+2017,Bordier+2022,Offner+2022,Bordier+2024,Frost+2025}. This has fundamental consequences on the final fate of the overall system, including the nature and properties of their compact remnants \citep{Perets+2012,Vigna-Gomez+2021,Vigna-Gomez+2022,Stegmann+2022}. Although close binaries are expected to interact strongly through mass transfer, triple systems can undergo long-term secular interactions that may alter their configuration until they either settle into a stable structure or become disrupted. Consequently, evolutionary models are increasingly incorporating both binary stellar evolution and multi-body dynamical effects \citep{Toonen+2016,Toonen+2020}. While the evolution of massive single and binary stars has been extensively studied, a new Pandora's box opens up for the evolutionary pathways of massive triples. 

In this work, we focused on hierarchical triple systems, characterised by an inner binary and a more distant tertiary companion. We specifically examined relatively tight configurations where all three components are within approximately 200~au of one another. In such systems, the inner binary may undergo standard binary interactions, while the outer companion can influence the secular evolution through a combination of dynamical effects and tidal interactions \citep{Tokovinin+2004,Naoz+2016,Toonen+2020,Kummer+2023}. 

One such dynamical effect is the von Zeipel-Kozai-Lidov (ZKL) mechanism \citep{Von-Zeipel+1910,Kozai1962,Lidov1962}, which involves the periodic exchange of eccentricity and relative inclination between the inner and outer orbits \citep[for reviews on ZKL, see][]{Naoz+2016,Libert2022}. These fluctuations occur on timescales exceeding the orbital periods and can lead to the tightening of the inner binary, companion exchange or, in some cases, stellar coalescence \citep{Michaely+2014,Toonen+2020}.

Recent population synthesis simulations have highlighted the impact of tertiary companions on massive star evolution. For instance, \citet{Kummer+2023} demonstrated that primary evolutionary channels such as mass transfer, orbital unbinding, or dynamical destabilisation are sensitive to the semi-major axes and mass ratios of the system. However, a significant source of uncertainty in these models remains the lack of observational constraints on the distributions of physical and orbital properties for these systems.

For this work we used main-sequence O-type triples observed in the Southern MAssive Stars at High angular resolution survey \citep[SMaSH+;][]{Sana+2014} to provide the first realistic observational distributions of key physical parameters, including the masses, mass ratios, and separation ratios. To perform this analysis, we adopted the inner orbital parameters derived by \citet{Tramper+2026}. We also investigated whether these systems satisfy the dynamical stability criterion of \citet{Mardling+2002} and evaluated the relevant dynamical timescales governing their evolution.

The paper is structured as follows. Section \ref{section:obs} describes the sample selection and the adopted physical parameters. In Sect. \ref{section:results} we present the results of our analysis, including the examination of correlations between system parameters, the introduction of the dynamical stability criteria, and the characterisation of both the marginal and joint distributions of masses, mass ratios, and separations. Section \ref{section:discussion} discusses the relevance of the characteristic dynamical timescales, evaluates the observational completeness and associated biases, and addresses the evolutionary implications of these systems. Finally, our conclusions are given in Sect. \ref{section:Conclusions}.

\section{Sample definition}
\label{section:obs}


\subsection{Selection criteria}
\label{observations}

The study of hierarchical triple systems requires an observational multi-technique approach, as their orbital properties span several orders of magnitude in separation, ranging from $\sim 10^{-2}$~au for compact inner binaries to $10^{2}$~au for wide outer companions. Tight inner binaries are typically identified through spectroscopic or photometric variability, while wider companions are resolved using high angular resolution techniques. 

Our sample is primarily drawn from the SMaSH+ survey, which targeted more than 150 Galactic main-sequence O-type stars and probed, among others, for close companionship in the 1–100 mas range \citep{Sana+2014}. To maximise the number of hierarchical triple systems within projected separations of $\lesssim$200~au (dictated by the outer working angle of the Very Large Telescope Interferometer Precision Integrated-Optics Near-infrared Imaging ExpeRiment \citep[VLTI/PIONIER, hereafter PIONIER;][]{PIONIER}, we applied a uniform set of selection criteria to detections from both PIONIER interferometric observations and Very Large Telescope (VLT) aperture-masking observations obtained with the Nasmyth Adaptive Optics System Near-Infrared Imager and Spectrograph \citep[NACO;][]{Rousset+2003} in sparse aperture masking (SAM) mode. Among the 73 objects observed with PIONIER and/or NACO/SAM, 52 exhibit at least one detected companion with a contrast of $\Delta H \leq 4$.

From this parent sample, we selected O-type hierarchical triple systems according to the following criteria:
\begin{enumerate}
    \item The system must be a hierarchical triple within $\rho\approx \tilde{a}_{\mathrm{out}}/d\leq100$~mas (corresponding to \aout\,$\lesssim200$~au; see Sect. \ref{subsec:parameters}), excluding higher-order configurations (e.g. 2+2 quadruples or more complex hierarchies). Systems with additional companions beyond 
    $\sim$1000~au (detected via direct imaging) are included, provided the inner triple remains within the $\sim$200~au limit.
    \item The outer companion must be spatially resolved, through interferometry (PIONIER) and/or aperture masking (NACO/SAM).
    \item The system must satisfy the contrast limit compatible with detection: $\Delta H \leq 4$ mag. 
\end{enumerate}

When a companion is detected by both PIONIER and NACO/SAM, the PIONIER parameters are preferred, as the higher angular resolution typically provides a more precise estimation of the projected separation with lower uncertainties.
We combined these detections with information from the literature to identify systems hosting an inner companion, while excluding cases where spectroscopic and high angular resolution observations refer to the same pair. To ensure the sample is not biased by overlooked low-mass eclipsing binaries that might elude radial velocity detection, we performed a photometric screening of all 73 sources using available NASA’s Transiting Exoplanet Survey Satellite (TESS) data and the automated classification algorithm of \citet{IJspeert+2024}. The results of this screening led to the exclusion of HD~93160 as a hierarchical quadruple, but resulted in no missing additions to the sample, as detailed in Appendix \ref{sec:TESS}. The detailed discussion of excluded or ambiguous systems is presented in Appendix \ref{appendix:triples?}. 

\begin{figure}
\centering
\includegraphics[scale=0.5]{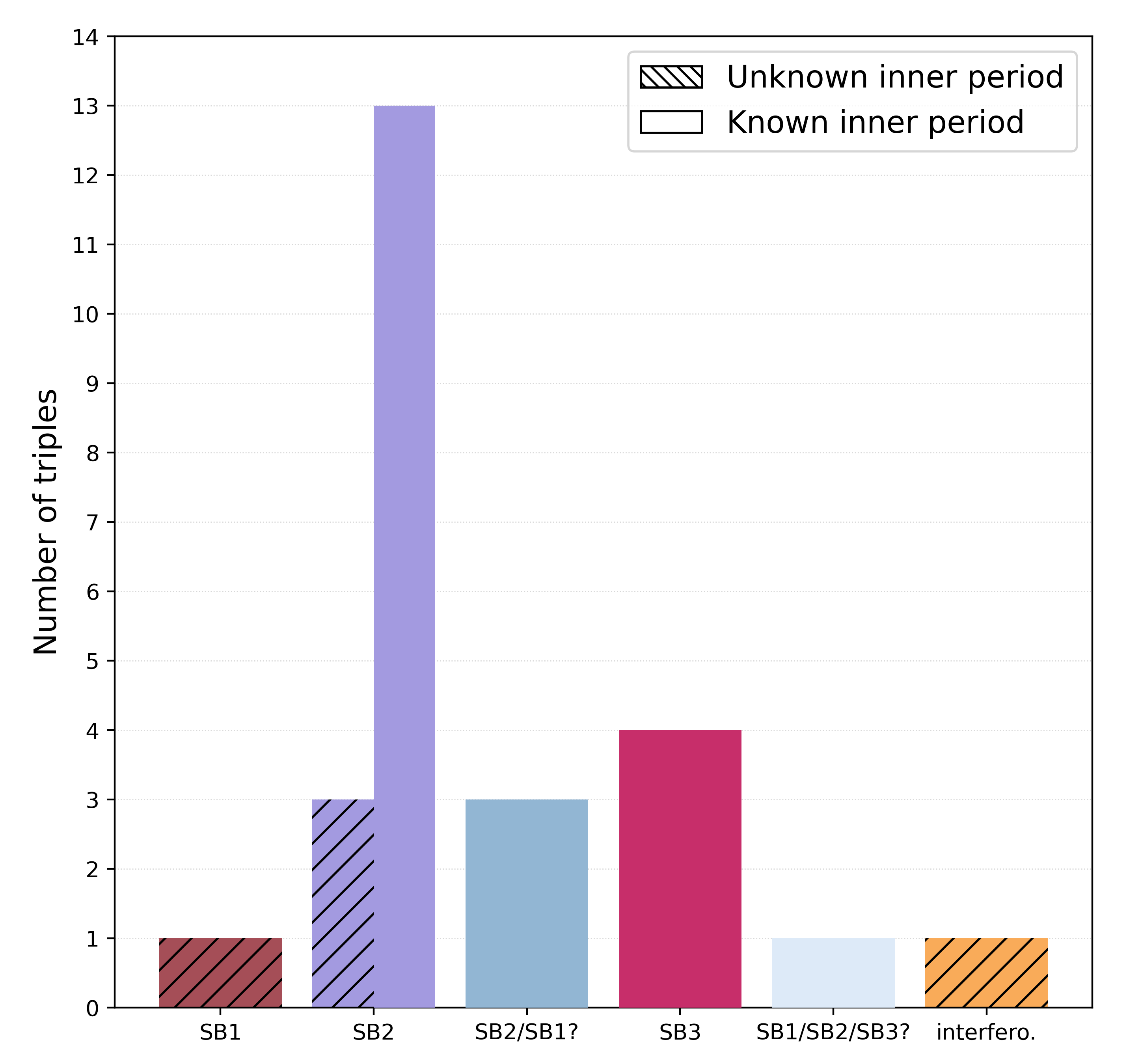}
\caption{Overview of the sample drawn from the SMaSH+ parent survey. The coloured areas refer to the inner binaries for which a measurement of the inner period ($P_{\text{in}}$) exists in the literature, while for the hatched areas a measurement of $P_{\text{in}}$ is not available to date. SB stands for spectroscopic binary and interfero stands for an interferometric triple system.}
\label{fig:description_sample}
\end{figure}

\begin{figure}
\centering
\includegraphics[scale=0.65]{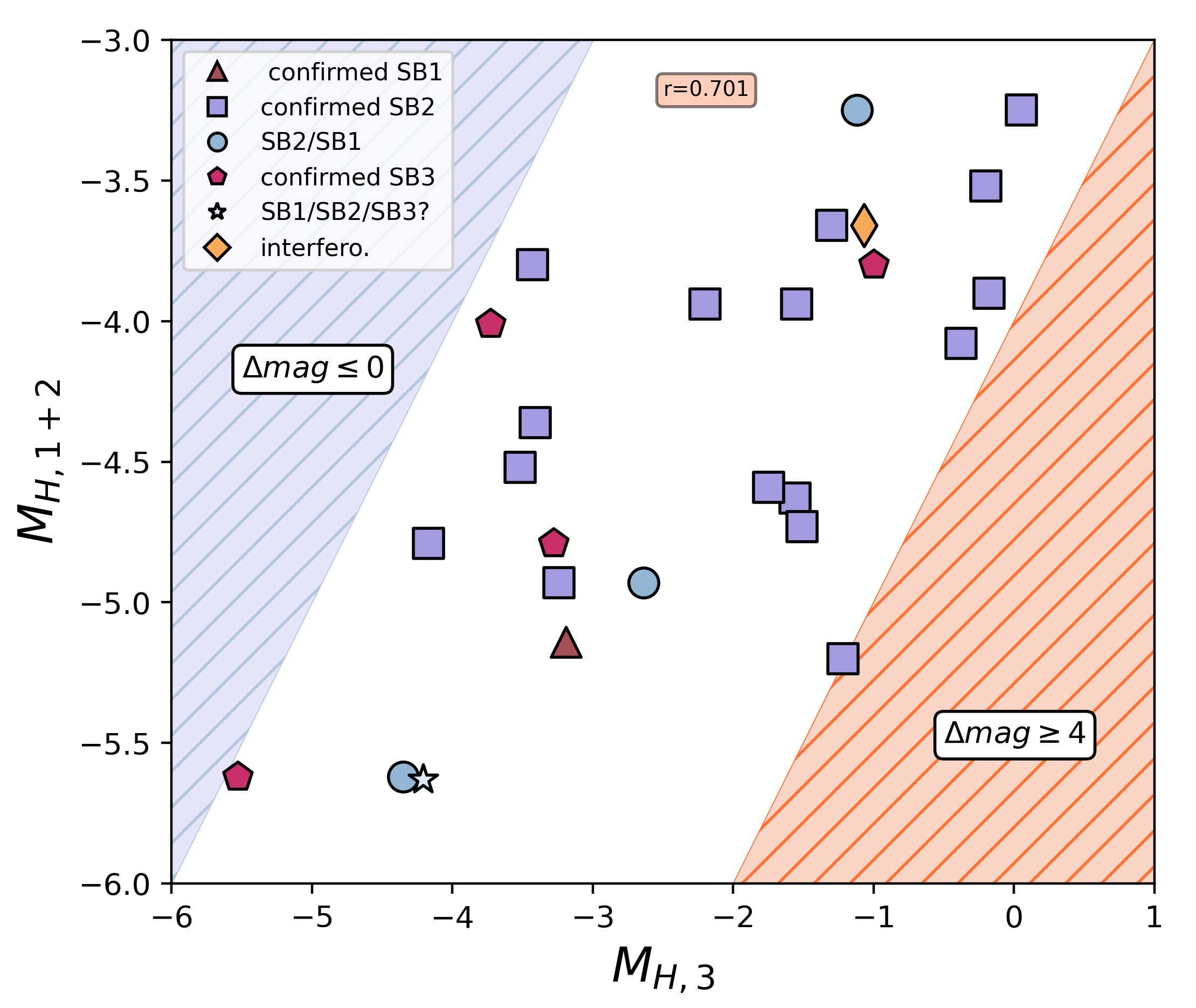}
\caption{Absolute $H$-band magnitudes of the inner and outer binaries within our sample. The red zone denotes the magnitude contrast inaccessible by PIONIER while the light purple area shows the magnitude difference greater than 0, meaning that any detected companion in that region would be brighter than the inner binary.}
\label{fig:magnitudes}
\end{figure}

Applying these criteria yields a final sample of 26 hierarchical triple systems, which forms the basis of the analysis presented in this work. Among them, 18 consist of an inner spectroscopic binary with a tertiary companion detected via PIONIER, while seven comprise an inner spectroscopic binary with a companion identified through aperture masking observations (NACO). Only one system, HD~101413, is a fully resolved triple, with both components detected through high angular resolution techniques (PIONIER and NACO). In total, 21 hierarchical triples have known orbital solutions for the inner pair. Detailed descriptions and parameters for each system in our sample are compiled in Appendix \ref{appendix:list_of_targets}. The full sample is presented in Fig. \ref{fig:description_sample}, while Fig. \ref{fig:magnitudes} illustrates the absolute $H$-band magnitudes of the inner and outer components. In the following, this set of hierarchical triples is referred to as the sample. 

While this study focuses primarily on hierarchical triple systems, it is worth noting that higher-order multiples (i.e. quadruple and higher-order systems) can, in many cases, be interpreted as hierarchical triples by considering the appropriate inner and outer pairs. The list of quadruple systems (within $\sim$1000~au) identified in the SMaSH+ dataset is provided in Appendix \ref{appendix:list_of_quadruples}. These systems are also illustrated in Figs. \ref{fig:m3_m1m2}, \ref{fig:m3_aout} and \ref{fig:stability_criterion}, where their parameters are overplotted in black for comparison. However, given the scope of the present work, we did not include a detailed analysis of these higher-order systems in the subsequent discussion of distributions or timescales.

By construction, all selected systems satisfy a brightness contrast of $\Delta H \le 4$, corresponding to the detection limit of PIONIER and translating into outer mass ratios of \qout $\gtrsim 0.11$ (using the calibrations of \citealt{Tramper+2026}). The explored parameter space is further defined by the angular resolution of the observations, with an inner working angle (IWA) of $\sim$1.5~mas and an outer working angle (OWA) of $\sim$75~mas. Since the targets span distances from approximately 0.8 to 3.5~kpc (median distance $\sim$1.8~kpc $\leftrightarrow a_{\rm OWA,50}\simeq136$~au; see Appendix \ref{appendix:list_of_targets}), these angular limits translate into a range of projected separations, with the physical OWA varying from $\sim$60 to $\sim$260 au across the sample. Consequently, the survey sensitivity does not exhibit a sharp cutoff in projected separation. For simplicity, we adopted a representative OWA of 200~au in the figures. This value corresponds approximately to the 89th percentile of the physical OWA distribution and therefore encompasses nearly 90\% of the targets in the sample, while closely reflecting the upper projected separation range effectively probed by PIONIER.

\subsection{Physical parameters and terminology}
\label{subsec:parameters}

\begin{figure}[!ht]
    \centering
    \includegraphics[width=1\linewidth]{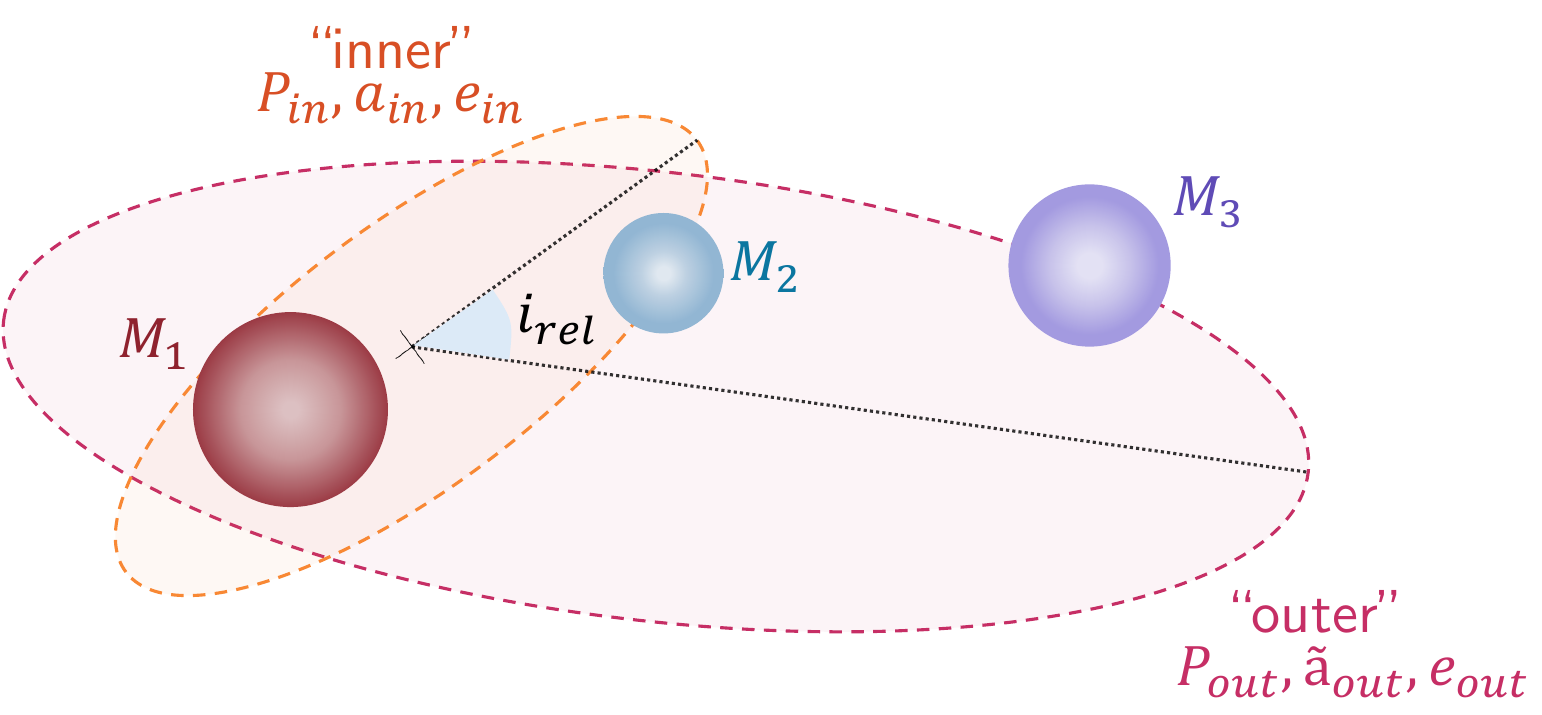}
    \caption{Sketch of a hierarchical triple system, composed of an inner tight binary and a third outer companion. The inner binary consists of two objects of masses $M_{\mathrm{1}}$ and $M_{\mathrm{2}}$, while the outer binary is formed by the pair of the inner binary and the third body of mass $M_{\mathrm{3}}$. The system configuration is not to scale. Inspired by \citet{Naoz+2013}.}
    \label{fig:sketch_triple}

\end{figure}

To characterise the architectural diversity of our sample, we adopted a hierarchical framework that decomposes each system into two nested orbital components. As illustrated in the sketch in Fig. \ref{fig:sketch_triple}, these systems consist of a tight inner binary (comprising masses $M_{1}$ and $M_{2}$) and a more distant tertiary companion ($M_{3}$). This outer component forms a second, wider orbit around the centre of mass of the inner pair. This standardised terminology allows us to consistently define key physical and orbital properties, such as mass ratios and separations, across a sample that ranges from compact spectroscopic pairs to widely resolved interferometric tertiaries.

 Most parameters referring to the inner and outer pairs are denoted with subscripts “in” and “out,” respectively. The total mass of the inner binary is  $M_{\mathrm{1+2}}$ = $M_{\mathrm{1}}$+$M_{\mathrm{2}}$ (with $M_{\mathrm{1}}$ the mass of the primary and $M_{\mathrm{2}}$ the mass of the secondary), while the tertiary mass is $M_{\mathrm{3}}$. Masses are derived from $H-$band magnitude contrasts following \citet{Tramper+2026}, who applied bolometric corrections and updated calibrations from \citet{Martins+2005} and \citet{Martins+2006} to determine mass-luminosity relations. Inner and outer masses are listed in Tables \ref{tab:sources_census_spectro} and \ref{tab:sources_census_interfero}. The outer mass ratio is defined as

\begin{equation}
    \centering
    q_{\text{out}}=\frac{M_{3}}{M_{1+2}}
.\end{equation}

Although \qout$> 1$ is uncommon ($<5\%$,\citealt{Tokovinin+2014b}), tertiary stars more massive than the inner binary have been observed \citep{Eisner+2022,Sanchez-Bermudez+2022, Vigna-Gomez+2022}. In an equal-mass triple, $M_{\mathrm{1}}=M_{\mathrm{2}}=M_{\mathrm{3}}$ leads to \qout$=0.5$ . 

Our analysis relies on single-epoch PIONIER (resp. NACO/SAM) observations, which provide six independent baselines (resp. one image) but prevent the reconstruction of full outer orbital solutions. Consequently, parameters such as outer eccentricities (\eout), arguments of pericentre, and mutual inclinations ($i_{\mathrm{rel}}$) remain unconstrained in this work. Given that the outer period cannot be reliably determined from a single snapshot, we maintained consistency by converting known inner periods ($P_{\mathrm{in}}$) taken from RV orbital solutions when available, into inner semi-major axes (\ain) using Kepler’s third law:\begin{equation}
    \centering
    a_{\mathrm{in}}^{3}=\frac{M_{\mathrm{1+2}}\,P_{\mathrm{in}}^{2}}{4\pi^{2}/G}.
    \label{eq:kepler}
\end{equation}

\noindent Here $G$ is the gravitational constant. 

The outer projected separation, \aout, is derived from the angular separation $\rho$ measured via PIONIER or NACO/SAM and the system distance $d$ as \aout $\approx d\rho$ (with $d$ in kpc and $\rho$ in mas). As this relies on single-epoch measurements, \aout\, represents an instantaneous projected separation rather than the true semi-major axis. We primarily adopted the spectrophotometric distances from \citet{Tramper+2026}, which were benchmarked against GAIA parallaxes, though alternative methods providing lower uncertainties were used for a small subset of systems. For more details, we refer  to   Sect. 2.1 of their paper. For our sample's typical distance of $\sim$2~kpc, the observed angular separations correspond to physical scales of approximately $\sim$3–200~au. As demonstrated by \citet{Tramper+2026}, for random orbital orientations and phases, corresponding to uniform priors on the cosine of the inclination ($\cos{i}$), the argument of periastron ($\omega$), and the eccentricity ($e$), the mode of the \aout\, distribution provides a reliable approximation of the true semi-major axis $a_{\mathrm{true}}$ to within approximately 10\%.

Tables \ref{tab:sources_census_spectro} and \ref{tab:sources_census_interfero} compile all collected and derived system parameters. Inner periods are taken from RV orbital solutions when available. Spectral classifications and binary types follow \citet{Sota+2014}, with updates noted where applicable.

\section{Results}
\label{section:results}
In this section we studied possible correlations among the physical parameters of the hierarchical systems: the mass of the inner and outer components ($M_{\text{1+2}}$ and $M_{\text{3}}$), the mass ratios (\qout), and their separations ($a_{\text{in}}$ and $\tilde{a}_{\text{out}}$).

\subsection{Observed stellar parameters: masses and separations} 
\label{subsect:mass_tertiary}

\begin{figure*}[!t]
\centering
\sidecaption
\subfigure[]{\includegraphics[width=6cm]{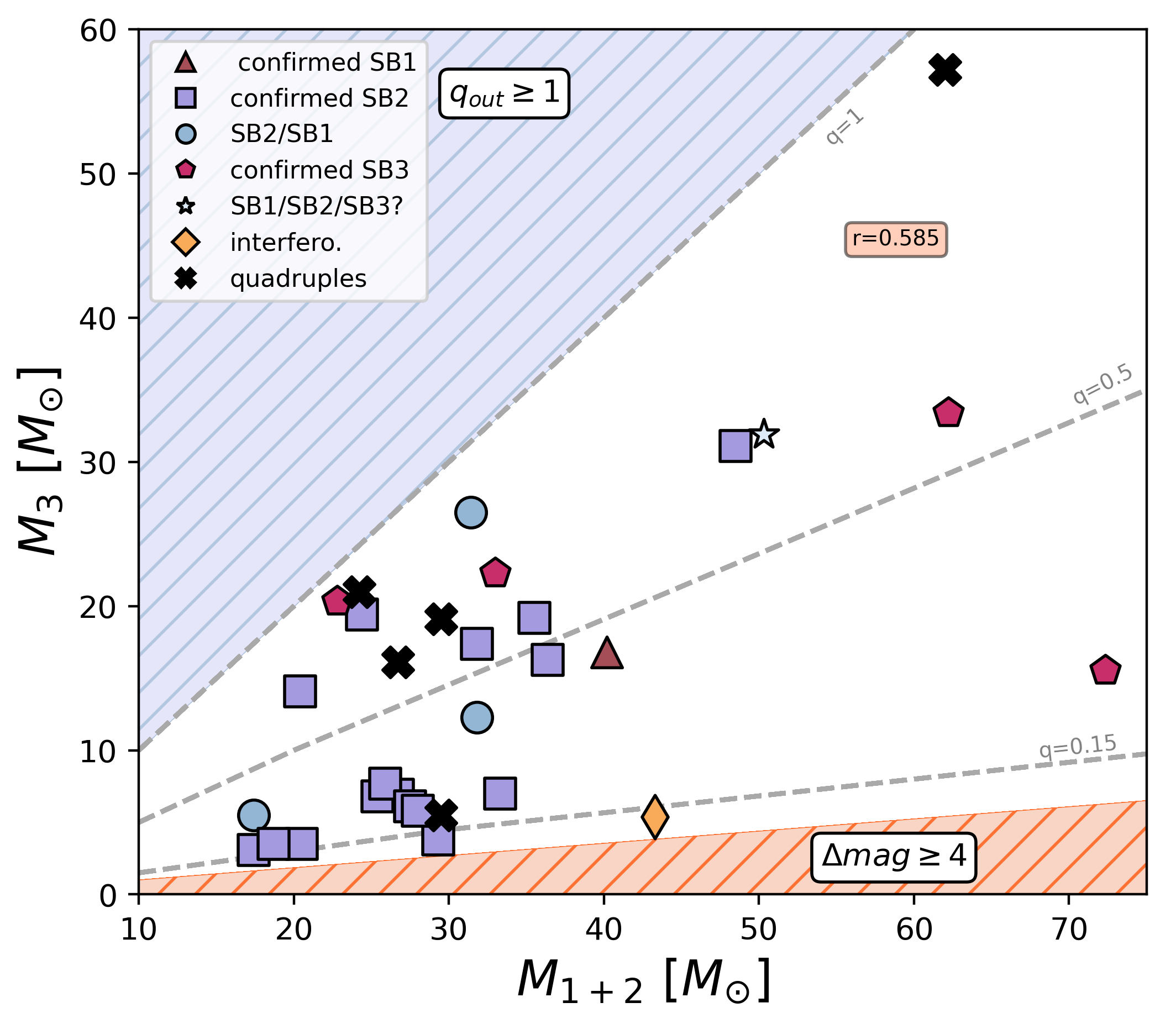}
    \label{fig:m3_m1m2}}
\subfigure[]{\includegraphics[width=6cm]{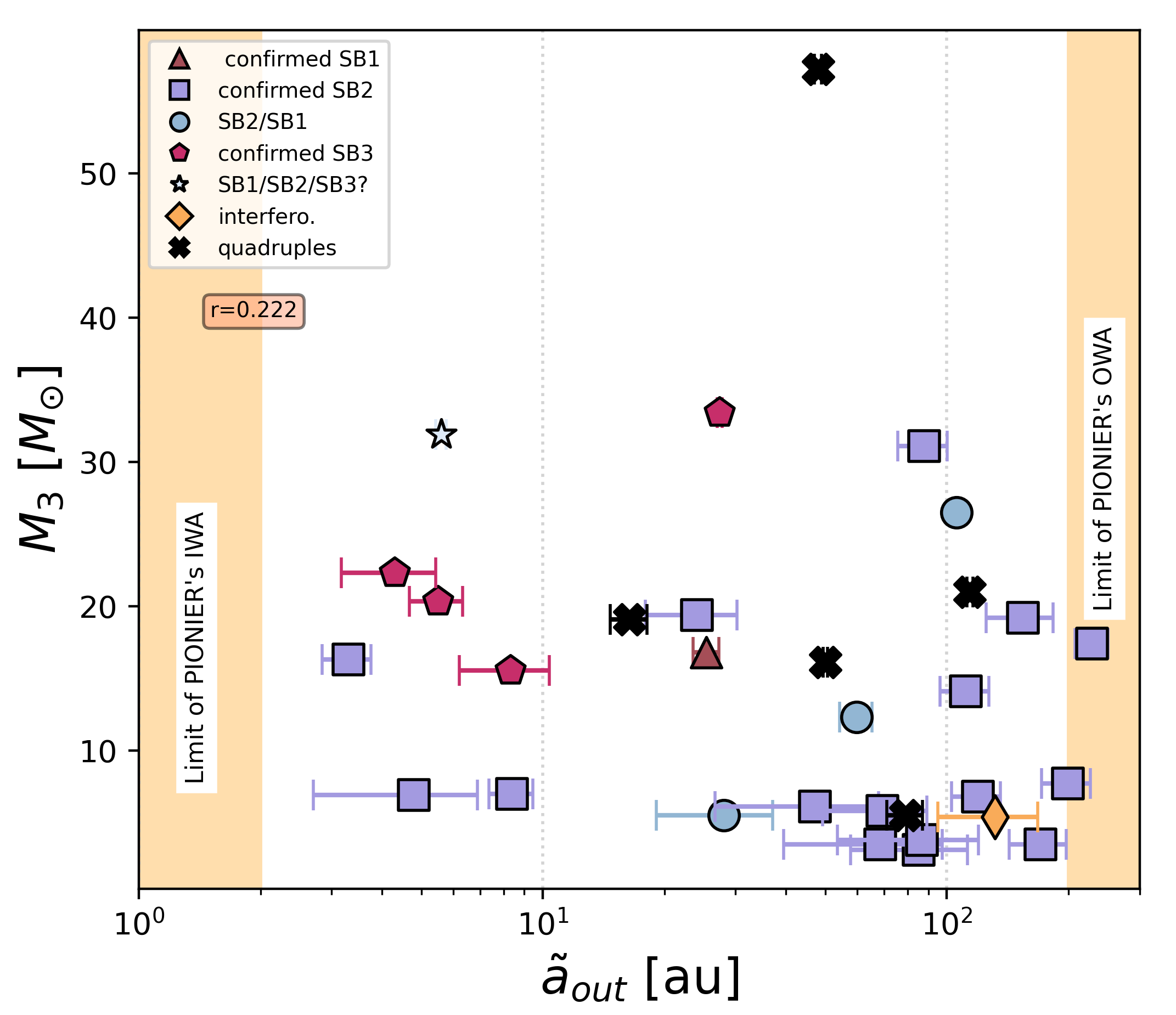}
    \label{fig:m3_aout}}
\caption{\textit{Panel a}: Mass of the outer companion ($M_{\mathrm{3}}$) with respect to the mass of the inner binary ($M_{\mathrm{1+2}}$). The coloured and hatched areas represent the regions where \qout$\geq1$ (blue) and $\Delta mag\leq4$ (orange), showing the contrast limit reachable by PIONIER. \textit{Panel b}: Mass of the tertiary ($M_{3}$) as a function of the outer separation \aout\, in au and logarithmic scale. The light orange shaded areas indicate the approximate PIONIER IWA and OWA limits. For both plots, the symbols indicate the type of the inner spectroscopic binary. The black crosses indicate the quadruples.}
\end{figure*}

In Fig. \ref{fig:m3_m1m2} we display the mass of the tertiary as a function of the inner binary mass. The correlation coefficient of 0.59 (with a $p$-value$<0.01$) indicates a moderate trend between the mass of the inner binary and the tertiary mass. More massive inner binaries ($M_{1+2}\geq40$~\Msun) tend to form or retain more massive tertiary companions ($M_{3}\geq30$~\Msun). In particular, HD~167971 stands out as one of the two most massive system of our sample with an inner binary of 62.2~\Msun\, and a tertiary with a mass of 33.4~\Msun \citep{Ibanoglu+2013,LeBouquin+2017}. However, observational limits (as indicated by the orange hatched lower detection boundary) may partially contribute to this apparent trend, as low-mass tertiaries are more difficult to detect around high-mass inner binaries (see Sect. \ref{sec:completeness}). 

In Fig. \ref{fig:m3_aout} we compare the mass of the tertiary with the corresponding outer separation. We do not notice a significant correlation between the mass of the tertiary and its separation (with a correlation coefficient of $\sim$0.2 at the 95\% confidence level). However, the lack of data points in the upper left corner of the figure shows that no massive companion ($M_{\mathrm{3}}>25$~\Msun) is found at low separation (\aout\,$\leq6$~au). PIONIER's sensitivity allows us to detect this range of massive companions with separation down to 1.5~mas. Consequently, any system with a tertiary more massive than 20~\Msun\, and in close proximity to its inner binary should be detectable. It is worth noting that for systems with highly eccentric orbits, the absence of detection in that region might result from the tertiary being in an unfavourable orbital phase at the time of the observation. However, this effect is largely compensated by the fact that companions in highly eccentric orbits spend the vast majority of their orbital period at large separations near apoastron, where they are more likely to be resolved, as illustrated by the phase-dependent separation ratios in Fig. 7 of \citet{Tramper+2026}. Hence, the overall probability of detecting companions remains rather uniform within the survey's sensitivity limits (see Sect. \ref{sec:completeness}), even when only a single epoch or snapshot is available. Hence, the observed lack of close-in, very massive tertiaries appears to be intrinsic rather than due to an observational bias. Conversely, at large separations (\aout\,$\geq50$~au) a large variety of tertiary masses is found, from 3.1 to 31.1~\Msun. No clear dependence is observed between the type of inner binary (indicated by different colours and symbols) and the properties of the outer companion. This is consistent with the expectation that the classification of the inner pair provides only a weak constraint on its physical nature.

\subsection{Inner and outer separations: stability criteria}
\label{subsect:inner_outer_sep}
In this subsection we verified the stability criteria of 22 hierarchical systems with well-determined orbital parameters, comprising 21 systems with known inner periods ($P_{in}$) and one system for which we adopted the projected separation (\ain) as a proxy for the semi-major axis.
We deliberately excluded the four systems for which spectroscopic orbital solutions are not yet available (see Appendix \ref{appendix:list_of_targets}).  We refer to Appendix \ref{app:dyn_criteria} for a detailed description of the theoretical and empirical stability criteria adopted in this work. In particular, we consider the criterion of \citet{Mardling+2002} and \citet{Tokovinin+2004}, which provide a minimum separation (or period) ratio required for dynamical stability by accounting for the masses, eccentricities, and in Mardling's expression, the mutual inclination of the system.

\begin{figure}[!ht]
\centering
\includegraphics[scale=0.38]{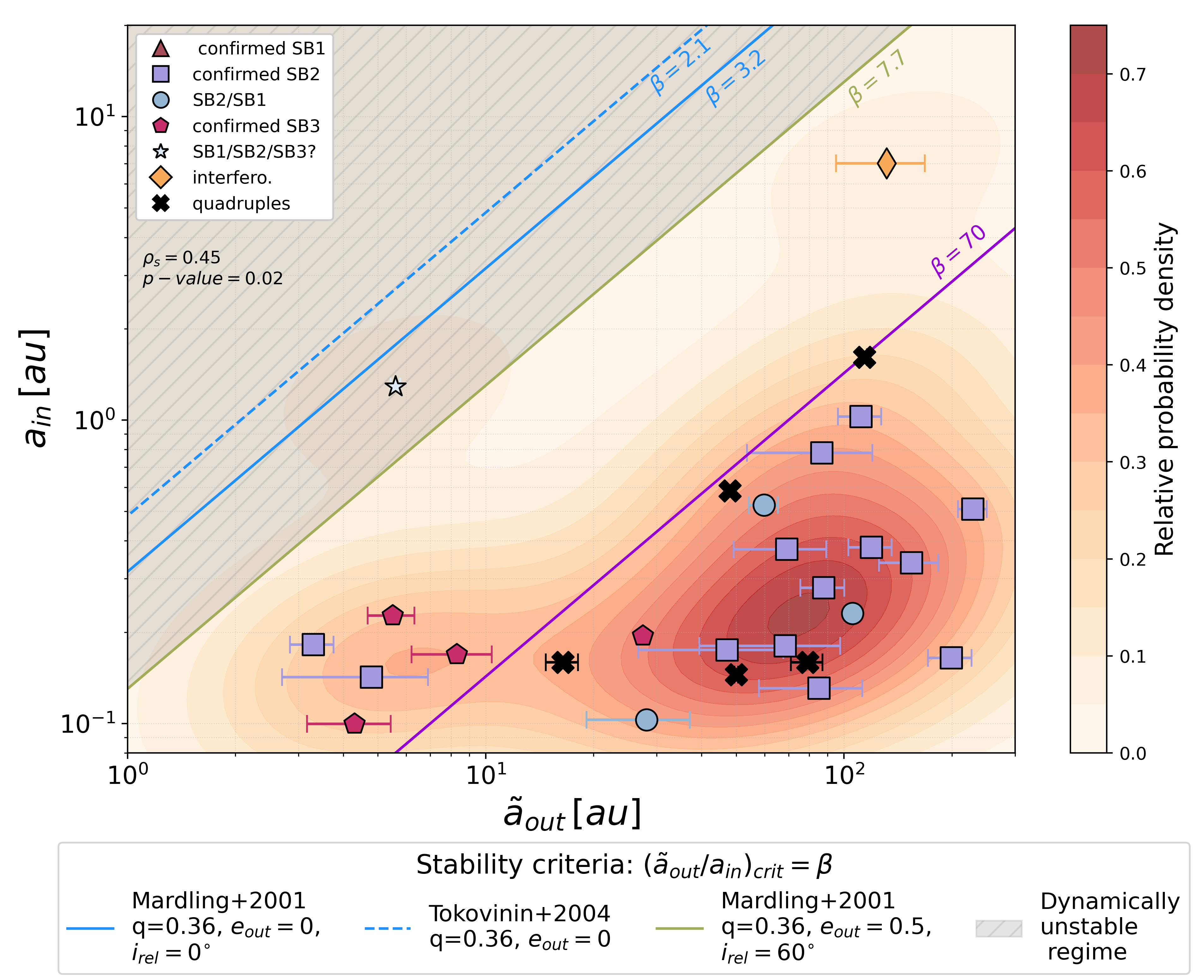}
\caption{Inner separation (\ain) as a function of the outer separation (\aout). Both quantities are expressed in au and are displayed on a logarithmic scale. The symbols indicate the type of inner spectroscopic binary. We also display the dynamically unstable regime (region in light grey) with reference values of $\beta$ for different mass ratios, eccentricities, and inclinations.}
\label{fig:stability_criterion}
\end{figure}

To quantify the overall architectural configuration of our sample, we analysed the hierarchy ratio, defined as the ratio of the outer projected separation to the inner semi-major axis ($\tilde{a}_{\mathrm{out}}/a_{\mathrm{in}}$). The distribution of this ratio confirms that our sample is composed of strongly hierarchical systems, with a median hierarchy ratio of 162.7 and a mean of 255.8. We found a statistically significant positive correlation between the inner and outer orbital scales, yielding a Spearman rank correlation coefficient of $\rho_{S}$=0.44 with a $p-$value of 0.04. This suggests that systems with wider inner binaries tend to host tertiaries at larger physical separations.

Figure \ref{fig:stability_criterion} displays the inner separation as a function of the outer projected separation, relative to different stability limits. For a better assessment, we overlaid the theoretical and empirical stability criteria (from \citealt{Mardling+2002} and \citealt{Tokovinin+2004}) with the corresponding $\beta$ factor (value of the critical separation ratio; see Eqs. \ref{eq:mardling_criteria} and \ref{eq:quasi_Tokovinin}). The limits were calculated using the median outer mass ratio of our sample \qout=0.36 (see Fig. \ref{fig:cdf_q}) with a varying outer eccentricity of 0 and 0.5. Increasing the outer eccentricity implies a larger separation ratio to guarantee the stability of the system, as $\frac{\tilde{a}_{\text{out}}}{a_{\text{in}}}$ is inversely proportional to $(1-$\eout)$^{2}$. Conversely, the mass ratio \qout\, has a lesser influence on the stability criterion.

The distribution in Fig. \ref{fig:stability_criterion} reveals two distinct architectural regimes: a more compact group characterised by both low \ain\, and low \aout, and a more decoupled population maintaining low \ain\, but extending to much larger \aout. Notably, the minimum observed hierarchy ratio in our sample is 4.34, which is close to the theoretical stability limit for typical mass ratios in O-type triples. The fact that all systems sit at or above this threshold indicates that our sample consists of long-term dynamically stable architectures, while effectively probing the edge of the stable parameter space. The majority of the triples (15 out of 22) and all the quadruples occupy the lower right region of the diagram with hierarchy ratios $\beta>70$. These systems are characterised by a very strong hierarchy, placing them deep within the stable dynamical regime and far from the critical limit. Five additional systems lie closer to the stability boundary in the lower left portion of the stable region, corresponding to relatively compact configurations with \ain$<$0.2~au and \aout$<$10~au. Interestingly, we found comparatively few stable systems with larger inner separations, with the exceptions of HD~101413, the only fully resolved triples in our sample. Detecting spectroscopic binaries with periods exceeding $\sim$1000~days remains challenging for radial velocity monitoring alone. However, interferometric observations effectively fill this gap. Given the capabilities of PIONIER, the scarcity of systems with large inner separations suggests they may be intrinsically uncommon, though residual observational biases cannot be entirely excluded (see Sect. \ref{sec:completeness}).

 For CPD$-47\degr2963$ (green star on Fig. \ref{fig:stability_criterion}), the present constraints, based on interferometric orbital parameters indicating a relatively high outer eccentricity \citep[\eout$=0.66$;][]{LeBouquin+2017} and a mass ratio of about 0.6, would instead suggest a stable configuration with a stability parameter $\beta$ closer to 10, compared to the inferred value of $\sim$4. This discrepancy likely reflects the significant uncertainties affecting this system. Further spectroscopic and interferometric observations are therefore required to clarify its nature, in particular to determine whether it represents a genuinely unstable configuration, or whether the high eccentricity leads to a misidentification of the interferometric companion with the spectroscopic one, in which case the system may in fact be a binary rather than a triple.

It is, however, important to keep in mind that the Mardling's stability limit is only an approximation and tends to be conservative. Systems near the stability edge would rather quickly be deemed unstable, leading to more false unstable than false stable systems. Dynamical stability should not be viewed as a strict yes-or-no state, but rather as a range defined over different timescales: systems deep within the unstable regime may dissolve rapidly, while those near or slightly beyond the critical limit can remain as such for long timespans. Numerical studies (low-mass primaries: \citealt{Toonen+2022}, high-mass primaries: \citealt{Bruenech+2024}) have shown that some systems deemed unstable by Mardling’s criterion can survive for more than a Hubble time. Considering that massive stars are short-lived, this may help explain the presence of apparently marginally stable systems in our sample.

\subsection{Distribution of masses, mass ratios and separations}
\label{sect:distribution_q_and_m}

In this section we derived analytical distributions representative of the observed properties of hierarchical triples in our sample. We also compared these distributions with well-known distributions derived for binary systems \citep[see e.g. ][]{Sana+2012,Moe+2017,Shenar+2022,Tramper+2026}. Historically, such distributions are often modelled by a power law, where the probability density function (PDF) is given by

\begin{equation}
    f(x_{i})\propto x_{i}^{\gamma_{i}}.
\end{equation}

Here $x_{i}$ represents the following quantities: $M_{\mathrm{1+2}}$,  $M_{\mathrm{3}}$, $M_{\mathrm{1+2+3}}$, \qout\, or $\log$(\aout) and $\gamma_{i}$ is the associated power-law index. In this formalism, $\gamma_{i}$=0 signifies a uniform probability distribution for the quantity $x_{i}$. We characterised the observed distributions by fitting power-law models to their cumulative distribution functions (CDFs).

\begin{figure*}[!ht]
\centering
\subfigure[]{\includegraphics[scale=0.42]{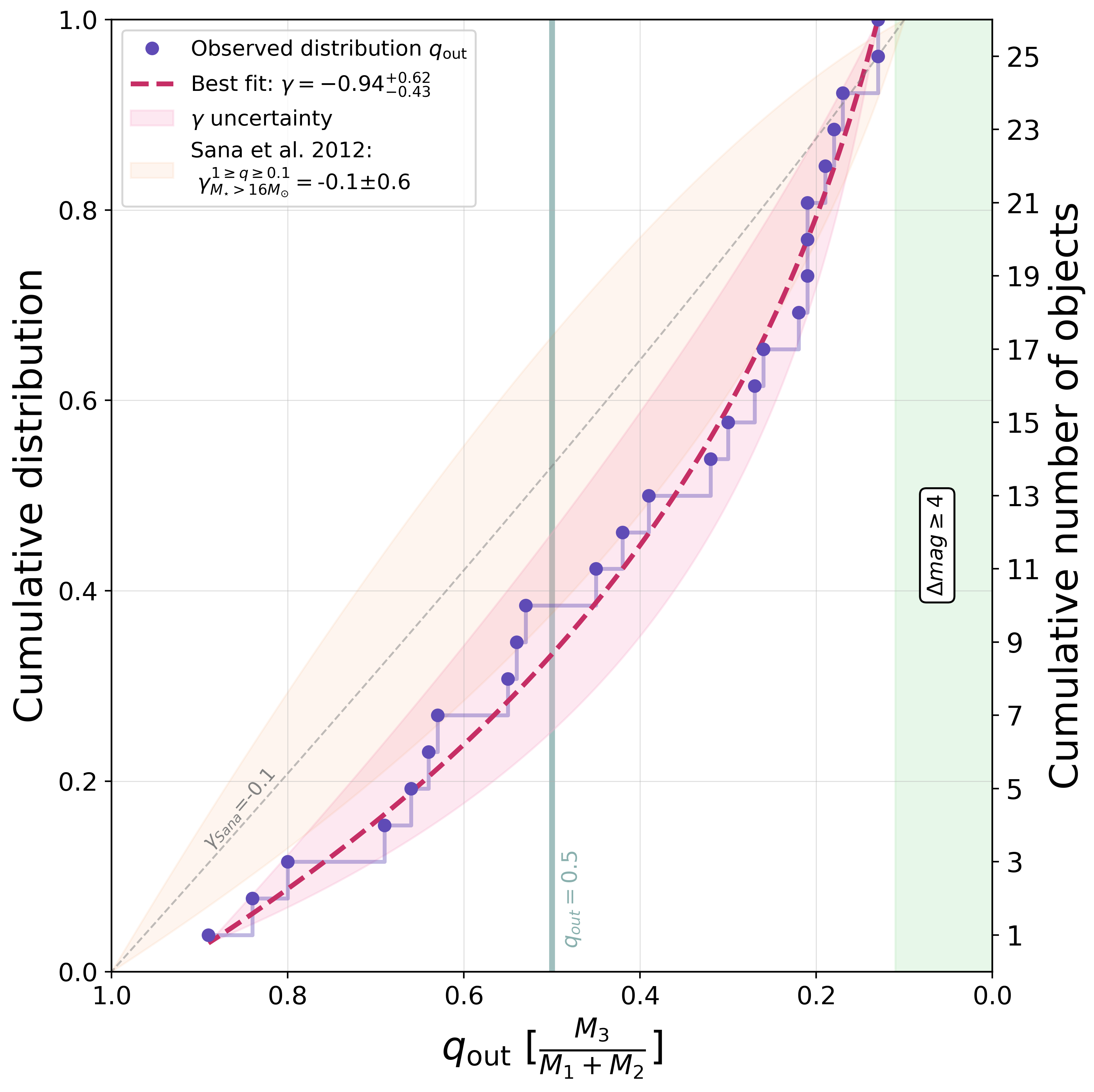}
\label{fig:cdf_q}}
\hspace{0.1cm}
\subfigure[]{\includegraphics[scale=0.42]{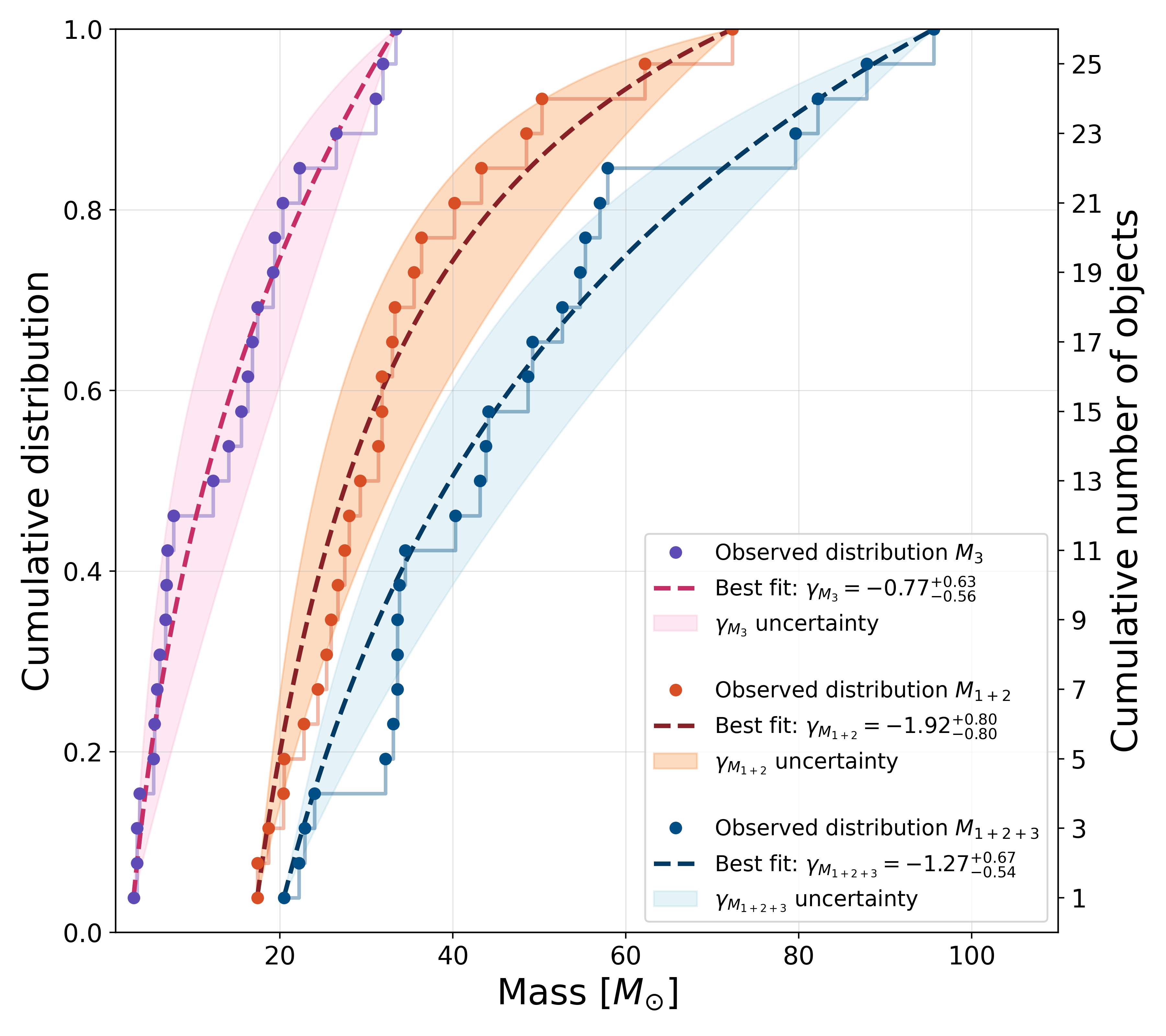} \label{fig:cdf_mass}}
\caption{\textit{Panel a}: Cumulative distributions of the outer mass ratio \qout\, (purple) of all detected companions. The dark pink line represents the best-fit power law;  the shaded area indicates its associated uncertainty. The light orange area denotes the power-law boundaries with $\gamma=-0.1\pm0.6$ as found by \citet{Sana+2012} for O-type binaries with a primary mass $\gtrsim$16~\Msun. The green area shows the mass ratios at which PIONIER's sensitivity drops. \textit{Panel b}: Cumulative distribution of $M_{3}$ (purple), $M_{1+2}$ (orange), and $M_{1+2+3}$ (blue) alongside best-fit power laws (dotted lines) and associated uncertainties (shaded regions).}
\end{figure*}

Figure \ref{fig:cdf_q} displays the cumulative distribution of mass ratios for our sample. The best-fit power law, with a central value of $\gamma_{\mathrm{q_{\mathrm{out}}}} = -0.94_{-0.43}^{+0.62}$, suggests a relatively flat distribution within the observed range, indicating no strong preference for very high or very low mass ratios within our detectable limits. This finding is broadly consistent with \citet{Sana+2012}, who reported a similar power-law exponent of $\gamma= -0.1\pm0.6$ for massive binaries with $M>$16~\Msun\, and mass ratios spanning values from 0.1 to 1.0 \citep[also see: ][]{Almeida+2017,Shenar+2022,Sana+2025}. Their study also found a uniform mass ratio distribution for binaries with periods up to approximately nine years, echoing our observation of no preference for equal-mass triples. Our analysis shows \qout$>$0.5 in 38$\pm$9\% of the systems, denoted by the \qout=0.5 mark in Fig. \ref{fig:cdf_q} (green line). This significant proportion underscores that the third component in nearly half of the systems is more massive than at least one of the two inner components, a characteristic that could potentially impact their evolutionary pathways, as further discussed in Sect. \ref{sec:evolution}. The apparent prevalence of high-mass tertiaries within 200~au is partly a result of our reduced sensitivity to lower-mass companions that would otherwise shift the distribution towards smaller mass ratios (see Sect. \ref{sec:completeness}).

Figure \ref{fig:cdf_mass} compares the cumulated distribution of observed masses of the inner binaries (orange), the tertiaries (purple) and the full system (blue). The distribution of inner masses follows a power law with $\gamma_{\mathrm{M_{1+2}}} = -1.92\pm0.80$. This significant negative component suggests a steeper decline in the number of systems with increasing total inner binary mass. Namely, lower total inner binary masses are more common than higher values within the observed sample. When fitting the distribution of masses of the O star primaries (16~\Msun\, $\leq$ M $\leq 50$~\Msun), \citet{Tramper+2026} found a lack of lower mass (M $<30$~\Msun) primaries compared to the observed companions that follow a Salpeter initial mass function. Their companion mass distribution is calculated without consideration for the hierarchical order of the system. It includes many wider companions, detected using imaging techniques at separations of up to $8000$~au using NACO. In contrast, the tertiary companion mass distribution observed in the present study spans masses from $\sim$3 to $\sim$33~\Msun\, and exhibits a flatter profile compared to the primary masses, with a best-fit powerlaw featuring $\gamma_{\mathrm{M_{3}}} = -0.77_{-0.56}^{+0.63}$. This power-law index suggests that tertiary companion masses are distributed nearly uniformly across the observed range, indicating that the mass of the outer component is not restricted to a specific mass interval and that hierarchical architectures accommodate a wide variety of companion masses around O-type inner binaries. The cumulative distribution of total system masses ($M_{\mathrm{1+2+3}}$) is well described by a power-law slope of $\gamma_{\mathrm{M_{1+2+3}}} = -1.27_{-0.54}^{+0.67}$. The total masses range from approximately $\sim$20 to $\sim$100~\Msun\, and most systems (70\%) have masses comprised between 30 and 60~\Msun. Two of the most massive systems, HD~150136 ($M_{\text{tot}}$=85.9~\Msun) and HD~167971 ($M_{\text{tot}}$=95.7~\Msun) possess well-established spectroscopic and interferometric orbital solutions and are therefore discussed further in the following section in the context of their dynamical stability.

\begin{figure}[!ht]
\centering
\includegraphics[scale=0.45]{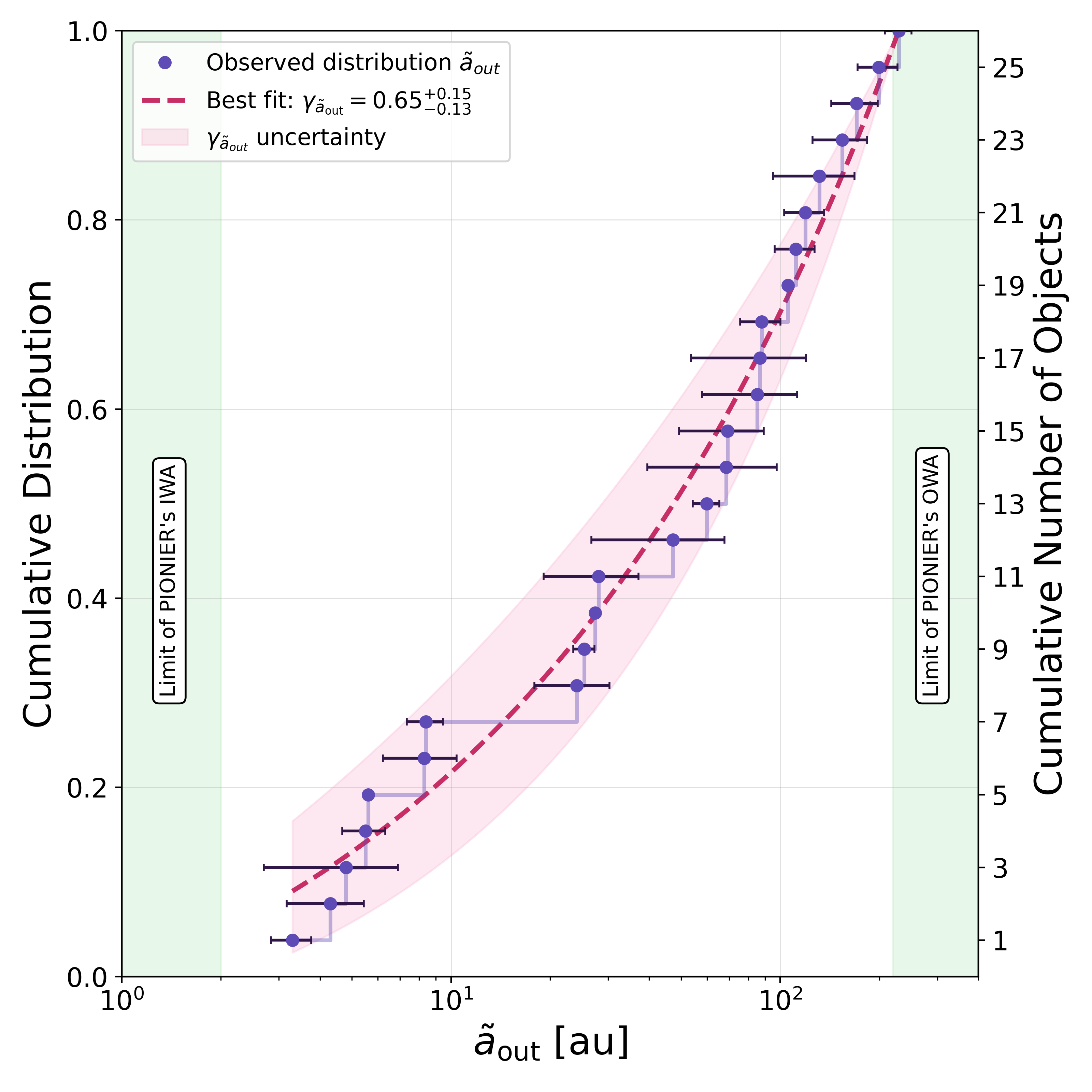}
\caption{Cumulative distribution of logarithmic outer separations for our sample. The best-fit power law is shown in dark pink;  its uncertainty is the pink shaded area. The green shaded areas indicate the approximate PIONIER IWA and OWA limits.}
\label{fig:sep_distribution}
\end{figure}

Figure \ref{fig:sep_distribution} illustrates the cumulative distribution of the outer projected separations. Unlike the results in \citet{Tramper+2026}, where the NACO in imaging mode allows detections near the $10^{4}$ separation range regime, the separations here only extend to the OWA of PIONIER (a few hundred  au), as chosen in the design of our sample (see Sect. \ref{section:obs}). More than half of the outer companions are found within 100~au. The cumulative distribution shows a slight flattening around 10-20~au. A rapid Kuiper's test gives $D_{K}=0.29$ with a $p$-value of 0.80  indicating that these deviations are statistically insignificant. Overall, the distribution shows a small but significant deviation from log-uniform, characterised by $\gamma_{\tilde{a}_{\mathrm{out}}} = 0.65_{-0.13}^{+0.15}$. Within the specific parameter space defined by our selection criteria, the sample remains representative and complete, although the inner and outer working angle limits of the instrument necessarily constrain detections at the boundaries of the sampled range (see Sect. \ref{sec:completeness}).

\subsection{Empirical joint distributions}
\label{sect:joint_distributions}

While the marginal distributions of separations and mass ratios are informative, the true architecture of hierarchical systems is best described by their joint distributions. In this section we present three empirical joint probability density functions derived from our sample of 26 O-type hierarchical triples.

To address the potential of this study as an input for population synthesis models, we investigated the joint PDFs, $f(x, y)$, for several combinations of fundamental parameters. Following the methodology of \citet{Moe+2017}, we defined the joint PDF $f(x, y)$ such that the probability $d^2P$ of finding a system within the infinitesimal range $[x, x + dx]$ and $[y, y + dy]$ is given by
\begin{equation}
    d^2P = f(x, y) \, dx \, dy.
\end{equation}
In cases where a parameter spans several orders of magnitude, such as orbital separations or hierarchy ratios, we performed our analysis in $\log_{10}$ space (i.e. $x = \log_{10} X$). 

A key question in the formation of multiple systems is whether these parameters are physically coupled or independent. If the formation and subsequent evolution follow an independent pairing scenario, where the two parameters are drawn independently from their respective parent distributions, the joint PDF should be separable into its marginal distributions, such that $f(x, y) \approx f_x(x) \cdot f_y(y)$. 

To test this hypothesis and quantify the degree of correlation, we calculated the Spearman’s rank correlation coefficient ($\rho_s$), a non-parametric measure widely used to assess the independence of orbital parameters in stellar multiples \citep[e.g.][]{Tokovinin+2014b, Smith+2024}. A correlation coefficient near zero, supported by a $p\text{-value} > 0.05$, allows us to conclude that the variables are statistically decoupled, justifying the use of independent marginal distributions in population synthesis frameworks. To visualise the joint probability density functions, we employed kernel density estimation (KDE), which smooths the discrete data points into continuous density contours. This non-parametric approach allows us to identify localised regions of high probability density without being constrained by fixed binning artefacts.

\begin{figure}[t]
    \centering
    \includegraphics[scale=0.45]{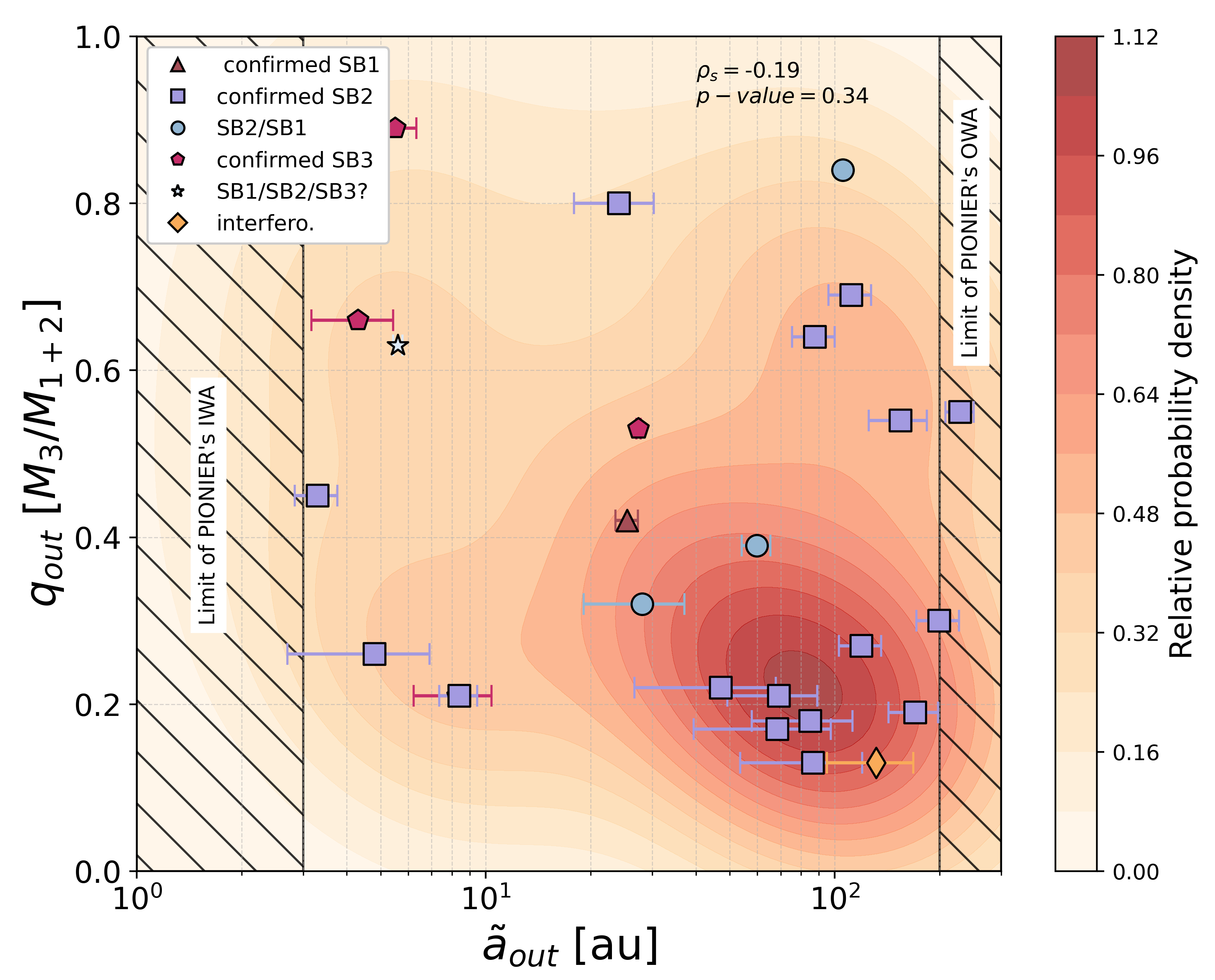}
    \caption{Joint PDF of outer separation $\tilde{a}_{\text{out}}$ and mass ratio $q_{\text{out}}$. The contours show the relative probability density and the different symbols indicate different type of inner binaries. For reference, the hatched regions represent the approximate PIONIER IWA and OWA limits.}
    \label{fig:joint_a_q}
\end{figure}

The joint distribution $f(\tilde{a}_{\text{out}}, q_{\text{out}})$ of the outer projected separation and the tertiary mass ratio is shown in Fig.~\ref{fig:joint_a_q}. We found a Spearman correlation coefficient of $\rho_{s} =-0.19$ with $p = 0.34$. The high $p$-value ($p > 0.05$) indicates that the instantaneous projected separation of the tertiary is statistically independent of its relative mass. While both the smallest (\qout$=0.13$) and one of the largest mass ratios (\qout$=0.84$) are found for systems with a projected outer separation of about 100~au, the highest outer mass ratio in the sample is actually observed at a closer separation of approximately 5~au. The primary density peak in our sample identifies a preferred architectural configuration, where tertiary companions with mass ratios of \qout$\approx$0.3 are most frequently distributed at projected separations between 30 and 100~au. The lack of a diagonal trend in the KDE contours suggests that the formation mechanism responsible for the tertiary, likely disk or turbulent fragmentation, does not result in specific mass ratios at specific distances within the 10-200~au range. Similarly, we investigated the relationship between the total system mass ($M_{1+2+3}$) and the outer separation (\aout). As shown in Fig.~\ref{fig:joint_mtot_a}, we found no significant correlation between these parameters ($\rho_{s}=-0.28,p=0.17$). This again tends to show that the separation of hierarchical tertiaries is decoupled from the total gravitational potential of the system.

\begin{figure}[t]
    \centering
    \includegraphics[scale=0.45]{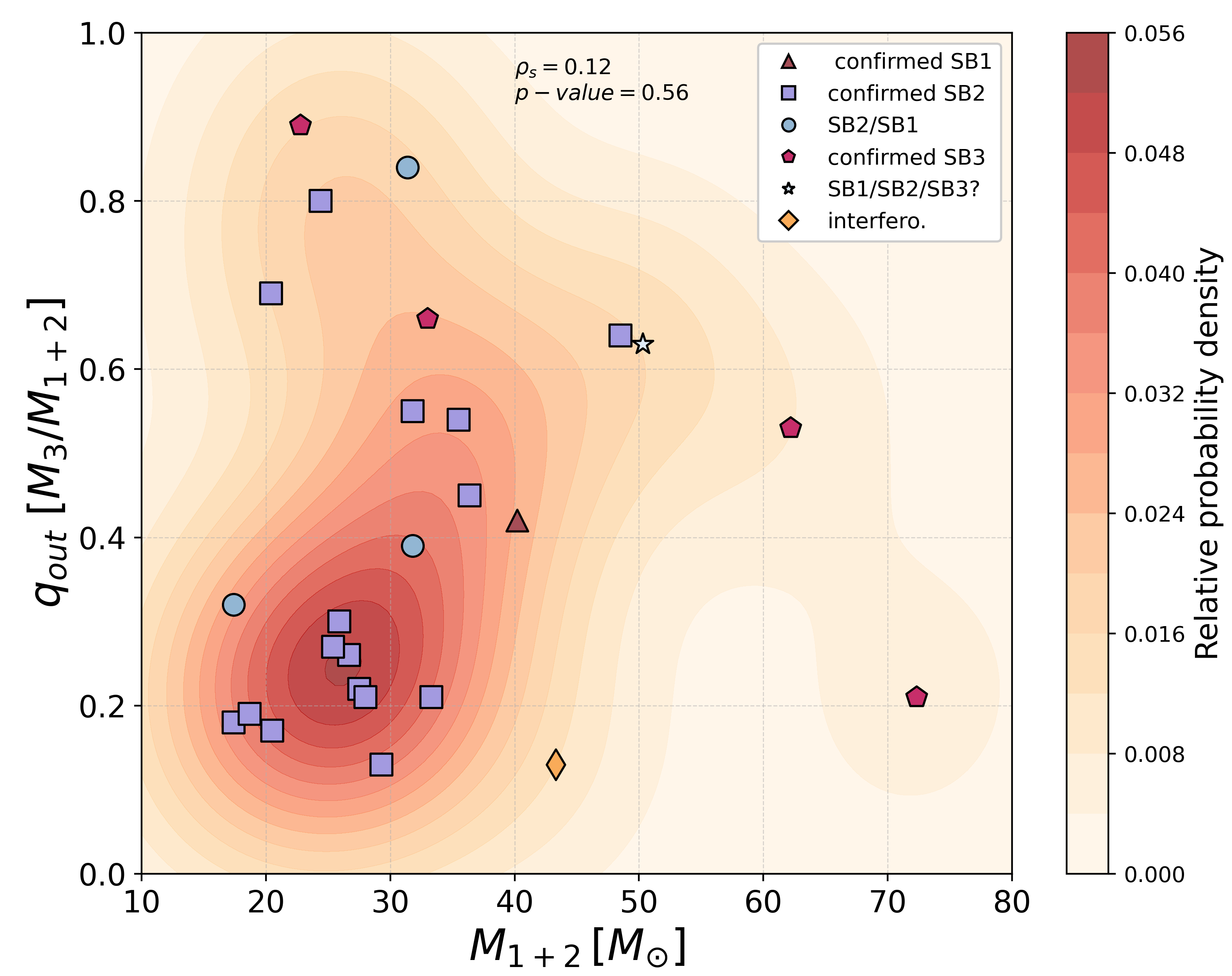}
    \caption{Joint PDF of primary mass $M_{1+2}$ and mass ratio $q_{\text{out}}$. The contours show the relative probability density, and the different symbols indicate the different types of inner binaries.}
    \label{fig:joint_mprim_q}
\end{figure}

The joint distribution $f(M_{1+2},q_{out})$ (Fig. \ref{fig:joint_mprim_q}) reveals a scale-invariant architecture for the outer components of O-type triples. The lack of a statistically significant correlation ($\rho_{s}=-0.12,p=0.56$) indicates that the relative mass of the tertiary is decoupled from the primary mass of the system. For relatively low-mass inner binaries ($M_{1+2}\leq35$~\Msun), we observed a wide range of mass ratios ($0.18 \leq$\qout$ \leq 0.86$), whereas higher-mass inner binaries exhibit a narrower range ($0.13 \leq $\qout$ \leq 0.62$). The density hotspot observed near $M_{1+2}\approx30$~\Msun\, likely reflects the intrinsic shape of the IMF within the O-type regime, where lower-mass O-stars are naturally more abundant. The critical finding is that the range and distribution of \qout\, remain constant as $M_{1+2}$ increases, suggesting that the tertiary formation mechanism is scale-invariant across the O-star mass spectrum. This supports the hypothesis that the relative mass of the third body is not dictated by the total mass reservoir of the parent core.

\begin{figure}[t]
    \centering
    \includegraphics[scale=0.4
    ]{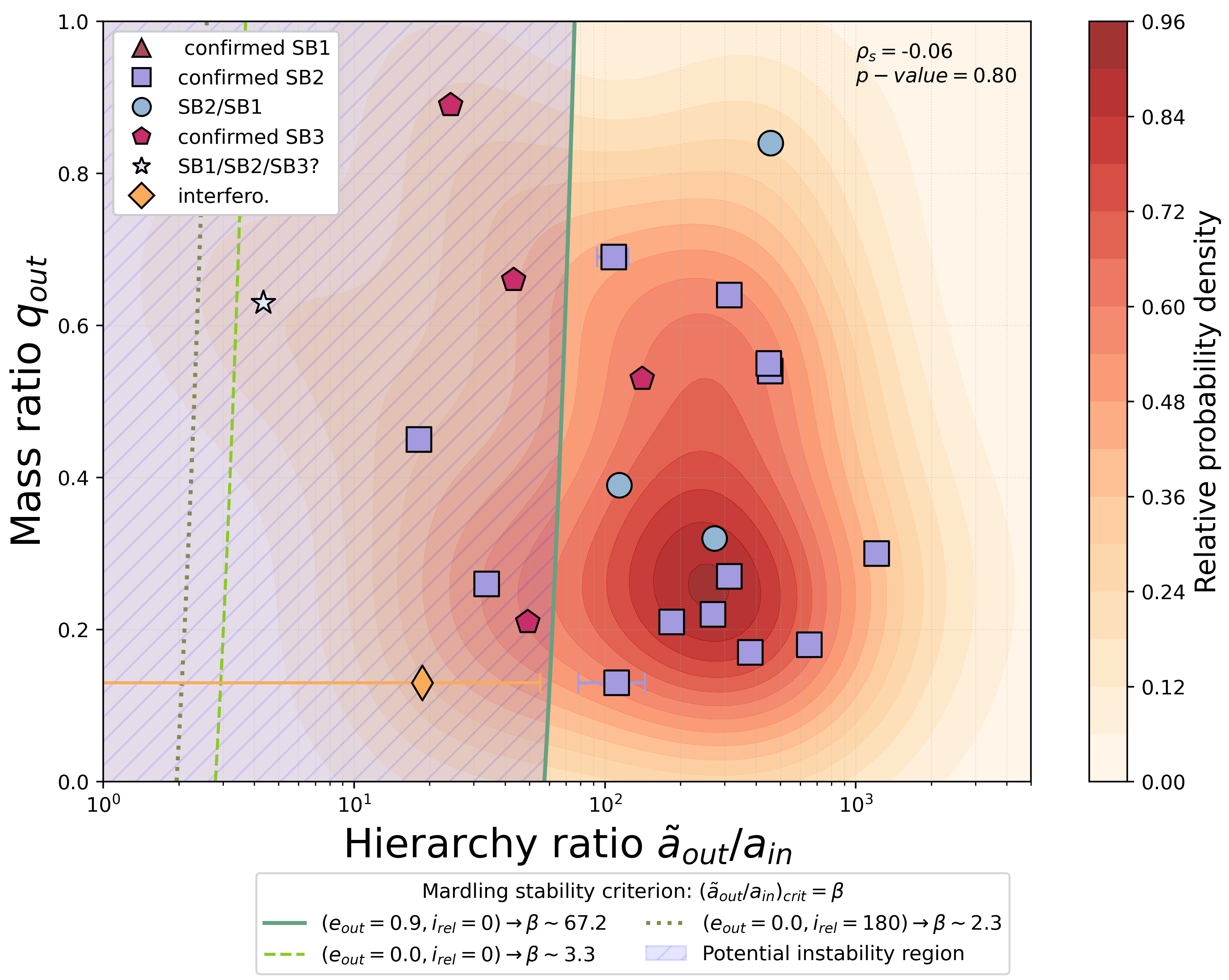}
    \caption{Joint PDF of hierarchy ratio ($\tilde{a}_{\text{out}}/a_{\text{in}}$) and mass ratio ($q_{\text{out}}$). The stability limits from \citet{Mardling+2002} are overlaid for eccentricities $e=0$ and $0.9$. The contours show the relative probability density, and the different symbols indicate the different types of inner binaries. }
    \label{fig:joint_ratio_q}
\end{figure}

The last model presented here for triple architecture is the joint distribution of the hierarchy ratio ($\tilde{a}_{\text{out}}/a_{\text{in}}$) and the mass ratio ($q_{\text{out}}$), presented in Fig.~\ref{fig:joint_ratio_q}. Here, the Spearman correlation is nearly zero ($\rho_s = -0.06$) with a $p$-value of $0.80$, representing a statistically null result. We overlaid the dynamical stability criteria from \citet{Mardling+2002} to delineate the physically allowed parameter space. We highlight three cases: the standard circular coplanar baseline ($e_{\text{out}}=0, i_{\text{rel}}=0^{\circ}$; green dashed), the least restrictive retrograde case ($e_{\text{out}}=0, i_{\text{rel}}=180^{\circ}$; green dotted), and the most restrictive eccentric case ($e_{\text{out}}=0.9, i_{\text{rel}}=0^{\circ}$; plain green). 

While a few targets appear to lie to the left of these boundaries (indicating an apparent instability), this is likely a consequence of the currently unconstrained $e_{\text{out}}$ and $i_{\text{rel}}$. A system appearing unstable in a prograde projection may be perfectly stable in a retrograde or more circular configuration. In the absence of a complete three-dimensional orbital solution, these systems may remain dynamically stable under orbital configurations not captured by our standardised baseline assumptions. However, the bulk of the probability density resides far to the right of the stability lines ($\tilde{a}_{\text{out}}/a_{\text{in}} > 70$), revealing a significant stability buffer. This suggests that massive hierarchical triples either originate in or dynamically relax towards configurations that are substantially more decoupled than the critical ratios required for long-term dynamical stability.

The independence observed across all three joint distributions ($\rho_s \approx 0$, $p-$value $\gg 0.05$) provides compelling evidence for an uncorrelated pairing for the outer companions in our sample of O-type triples. If the mass and orbit of the tertiary were coupled, we would expect to see a tilt in the KDE contours; instead, we found a broad, boxy distribution of probabilities.

Despite the sample size limitations, this empirical framework provides a robust starting point for improving population synthesis models. Our findings suggest that, to first order, synthetic populations of massive triples can be generated by sampling a simplified modelling approach:
\begin{enumerate}
    \item Sample $M_{\text{1+2}}$ from the IMF as suggested by \citet{Moe+2017} and $a_{\text{in}}$ from binary distribution functions.
    \item Independently sample $q_{\text{out}}$ and $\tilde{a}_{\text{out}}$ from the above marginal distributions, knowing that the projected separation \aout\, serves as a reliable proxy for the true outer semi-major axis $a_{\mathrm{true}}$ \citep{Tramper+2026}.
    \item Apply a dynamical stability filter: while a ratio $\tilde{a}_{\text{out}}/a_{\text{in}} > 5$ is often considered a conservative lower bound for stability, our study demonstrates that the population is well represented by significantly stronger hierarchies, typically exceeding $\tilde{a}_{\text{out}}/a_{\text{in}} > 70$.
\end{enumerate}

\section{Discussion}
\label{section:discussion}

Close binaries with very short orbital periods of just a few days, such as the ones of our sample (see Appendix \ref{appendix:list_of_targets}), are often considered the result of a ZKL migration due to a distant and highly inclined stellar companion. In this section we formulate the secular perturbations for hierarchical, non-coplanar triple systems within the ZKL framework and discuss the associated timescales for five systems with well-defined orbital solutions. Following this dynamical analysis, we evaluated the overall observational completeness of the SMaSH+ survey to determine how representative these configurations are of the broader massive triple population. By quantifying the spectroscopic and interferometric detection limits, we assessed the impact of observational biases on our results. Finally, we provide hints on the potential evolutionary pathways of these systems, discussing how the presence of the outer companion may influence the future stability and interaction history of the inner binary as the components evolve off the main sequence.

\subsection{Relevance of the ZKL timescales}
\label{subsect:KL_timescales}

The mutual inclination $i_{\text{rel}}$ between the two orbital planes of the inner and outer binaries (see Fig.~\ref{fig:sketch_triple}) is a priori not known for most systems. If we assume a significant mutual inclination, it is possible for the inner eccentricity to be induced to a high value in a cyclic variation with the mutual inclination. This dynamical evolution, originally described in the restricted three-body problem (e.g. for an asteroid in the Sun-Jupiter system), is referred to as the ZKL resonance \citep{Von-Zeipel+1910,Kozai1962,Lidov1962}. In recent years, numerous studies have demonstrated that ZKL oscillations may significantly influence the formation and evolution of various astrophysical systems \citep[e.g.][]{Shappee+2013,Fabrycky+2007,Toonen+2016,Bataille+2019,Vigna-Gomez+2022,Kummer+2023}.

We extracted a subsample of five systems from our triple sample, each with well-constrained inner and outer orbital parameters. These systems are listed in Table \ref{tab:kl_parameters}.
To assess the long-term dynamical stability of these configurations, we calculated the critical separation ratio, $\beta_{\text{crit}}$, according to the \citet{Mardling+2002} criterion (see Appendix \ref{app:dyn_criteria}), and compared it to the observed ratio, $\beta_{\text{obs}} = \tilde{a}_{\text{out}}/a_{\text{in}}$. As shown in Table~\ref{tab:kl_timescales}, four systems satisfy the condition $\beta_{\text{obs}} > \beta_{\text{crit}}$, confirming that they are currently in stable hierarchical arrangements. However, the margin of stability varies significantly across the subsample. The exception is \object{HD~152246}, where $\beta_{\text{obs}} < \beta_{\text{crit}}$, suggesting that this system may be in a marginally unstable or dynamically active state, potentially undergoing significant orbital evolution.

The timescale of the ZKL oscillations, which is much larger than the orbital period, is mostly sensitive to P$^{2}_{\text{out}}/\mathrm{P}_{\text{in}}$ and \eout. It was derived by \citet{Kiseleva+1998} and is expressed as 
\begin{equation}
    \tau_{\text{ZKL}}= \frac{2}{3\pi} \left(\frac{\mathrm{M}_{\text{1+2}}}{\mathrm{M}_{\text{3}}}+1\right) \left(\frac{\mathrm{P}^{2}_{\text{out}}}{\mathrm{P}_{\text{in}}}\right) \left(1-e^{2}_{\text{out}}\right)^{3/2}.
    \label{eq:ZKL_timescales}
\end{equation}

Table \ref{tab:kl_timescales} shows the timescale of the ZKL oscillations for each of the five systems. Overall, we note that the values of $\tau_{\text{KL}}$ are relatively low compared to those typically reported in the literature, where oscillation timescales are on the order of 1~Myr \citep{Antognini+2015,Kummer+2023}. This discrepancy arises from the considerably shorter outer periods ($\lesssim$20~years) contrasting with orbits of the order of thousand years in the aforementioned papers.

 \begin{table*}[!h]
    \centering
    \caption{Inner and outer binary parameters for five triples. }
    \begin{tabular}{ccccccccccccc}
    \hline
    \hline \\ [-1.5ex]
        & \multicolumn{5}{c}{Inner binary} & \multicolumn{5}{c}{Outer Binary} &  \\ 
        
        \cmidrule(r){2-6} \cmidrule(l){7-11} 
        Target & \ain & P$_{\text{in}}$ & $M_{\text{1+2}}$ ($M_{1}+\mathrm{M}_{2}$) & $e_{\mathrm{in}}$ & $i_{\text{in}}$  & \aout & P$_{\text{out}}$ & $M_{3}$ & \eout & $i_{\text{out}}$ & Ref.  \\
         & (au) & (days) & (\Msun) &  & ($\degr$)  & (au) & (days) & (\Msun) &  &($\degr$) &   \\
        \hline\\ [-1.5ex]
        
        HD~150136 & 0.17 &  2.67 & (42.8+29.5) &  - & 62.4 & 8.3 & 3144 & 15.5 & 0.682 & 106.11 & 1,2,3  \\

        HD~167971 & 0.19 & 3.3 & (32.3+30) & - & 73.76 &  27.4  & 7806 & 33.4 & 0.443 & 145.4 & 3,4 \\

        HD~135240 & 0.18 & 3.9 & (23.36+13.00) &  0.06 & 78.7 & 10.054  & 1603.24 & 16.29 & 0.529 & 78.171 & 5,6,7 \\

         HD~152246 & 0.23 & 6.00 & (22.8) & 0.09 & 30 & 5.5 & 470.7 & 20.4 & 0.843 & 112.46 & 8 \\

        HD~164740 & 0.1 & 1.54  & (20.5+12.5) & 0.0 & 52 & 3.9 & 496.82  & 22.3 & 0.219 & 75.3 & 9,10 \\

        \hline
         & 
    \end{tabular}

\tablefoot{These systems feature well-studied spectroscopic and interferometric orbital solutions. \aout\, are taken from the SMaSH+ survey when they are not given in the spectroscopic solutions.} 

    \tablebib{(1)~\citet{Mahy+2018}, (2)~\citet{Sana+2013}, (3)~\citet{LeBouquin+2017}, (4)~\citet{Ibanoglu+2013}, (5)~\citet{Mayer+2014}, (6)~Keskar et al. (in prep)., (7)~\citet{Svrckova+2026}, (8)~\citet{Nasseri+2014}, (9)~\citet{Sanchez-Bermudez+2022}, (10)~\citet{Campillay+2019} }
    
    \label{tab:kl_parameters}
    
 \end{table*}
 
The ZKL oscillations continue only as long as the perturbations from the outer companion govern the apsidal precession of the inner binary. Other physical processes, most notably general relativistic (GR) precession and distortions of the inner orbits due to tides and intrinsic stellar rotation, also induce apsidal precession and can effectively suppress ZKL oscillations if their associated timescales are shorter than $\tau_{\text{ZKL}}$ \citep[e.g.][]{Fabrycky+2007, Liu+2015, Toonen+2016, Bataille+2019, Lim+2020}.

As the five triple systems are not yet fully characterised, for instance, the radii of the components and their spin properties remain unknown, certain perturbations, such as tidal effects and rotational contributions in the inner binary cannot be calculated. 
For instance, the ratio of the ZKL timescale to the tidal effects timescale for apsidal precession on the inner binary,  which can be approximated as  \citep{Mardling+2002}

\begin{equation}
    \frac{\tau_{\text{ZKL}}}{\tau_{\text{\text{tide}}}}\sim 15 k_1 \frac{\mathrm{M}_2}{\mathrm{M}_1} \left(\frac{\mathrm{P}_\text{out}}{\mathrm{P}_\text{in}}\right)^2\left(\frac{1+q_\text{out}}{q_\text{out}} \right) \frac{(1-e_{\text{out}}^2)^{3/2}}{(1-e_{\text{in}}^2)^{5}} \left(\frac{r_1}{a_{\text{in}}}\right)^5,
    \label{eq:ZKL_timescales}
\end{equation}

\noindent  depends on the unknown values of the classical apsidal motion constant $k_1$ and the radius of the inner star $r_1$. Further characterisation of the triple systems will allow the significance of tidal effects to be assessed.

We therefore restricted our comparison to relativistic precession that can be derived using the known parameters. Although GR effects can typically be neglected for wide, interferometric-scale orbits, they may become significant in tight, short-period (day-scale) orbits. In these compact configurations, GR primarily impacts the inner companion's motion by advancing its pericentre. 
The equation of the precession induced by GR effects on the inner binary reads
\begin{equation}
    \dot{\omega}_{\mathrm{GR}}=\frac{3(G\mathrm{M}_{1+2})^{3/2}}{c^{2}a_{\mathrm{in}}^{5/2}(1-e_{\mathrm{in}}^{2})},
\end{equation}
with $c$ the speed of light, and the corresponding timescale is then $\tau_{\mathrm{GR}}=\frac{2\pi}{\dot{\omega}_{\mathrm{GR}}}$. When the inner eccentricity is not specified in the literature, we calculated the GR timescale using a minimum eccentricity of 0 and a maximum of 0.9. This approach provides a range of values and an estimate of the expected order of magnitude for these timescales.
The timescales are listed in Table \ref{tab:kl_timescales}. We found a dichotomy in the dominance of precession sources. For HD~150136 and HD~167971, the ZKL timescales are larger than those associated with GR effects. This indicates that, in their current orbital configurations, the relativistic precession detunes the secular resonance, effectively suppressing any high-amplitude eccentricity oscillations that the tertiary might otherwise induce in the inner binary. Unlike the aforementioned targets, the other systems in this subsample, 
HD~135240, HD~152246, and HD~164740, exhibit a different dynamical balance where the ZKL mechanism could potentially compete more effectively with other precession sources.

The ZKL mechanism may have played a critical role in the historical evolution of these systems, particularly in explaining the observed excess of short-period binaries. The presence of distant tertiary companions with high mutual inclinations has been proposed as a primary driver for the inward migration of the inner binary, ultimately leading to the tight inner pairs identified in our sample \citep[e.g.][]{Fabrycky+2007,Naoz+2014,Anderson+2017,Moe+2018,Bataille+2018}. However, the conditions for the ZKL-driven migration are demanding \citep[see][]{Bataille+2019} and it remains to be demonstrated that such conditions (e.g. high mutual inclination, strong dynamical coupling, minimal relativistic precession) can be encountered during the formation of triple star systems. Obtaining a complete three-dimensional characterisation of the  orbital parameters for our full sample is required to definitively address this question. 

\begin{table}[!h]
    \centering
    \caption{Dynamical parameters and characteristic timescales.} 
    \begin{tabular}{ccccc}
    \hline 
    \hline \\ [-1.5ex]
    Target & $\beta_{\text{crit}}$ & $\beta_{\text{obs}}$ & $\tau_{\text{ZKL}}$ & $\tau_{\text{GR}}$ \\
     & & & (yr) & (yr) \\
    \hline\\[-1.5ex]

    HD~150136 & 13.7  & 48.8  & 4766 & $[124-655]$ \\

    HD~167971 & 6.8 & 144.2 & 22149 & $[2.1-10.8]\times10^{2}$ \\

    HD~135240 & 9.5 & 55.9 & 756  &  2109 \\

    HD~152246 & 36.7 & 23.9 & 7.1  &  7805 \\

    HD~164740 & 4.8 & 39.0 & 214  &  563 \\

     \hline
     
    \label{tab:kl_timescales}

\end{tabular}

\tablefoot{Listed are the critical stability ratio ($\beta_{\text{crit}}$) calculated following \citet{Mardling+2002}, the observed separation ratio ($\beta_{\text{obs}}$), and the calculated timescales for the ZKL oscillations ($\tau_{\text{ZKL}}$) and general relativistic precession $\tau_{\text{GR}}$. When the inner eccentricity is not available in the literature, the general relativity timescale is estimated assuming two scenarios: a circular inner orbit, providing an upper limit on the timescale, and a highly eccentric orbit ($e_{\text{in}}=0.9$), representing a lower limit.}
\end{table}

\subsection{Observational biases and sample completeness}
\label{sec:completeness}

To address the potential biases in our sample and evaluate the statistical significance of our findings, we performed a quantitative completeness analysis of the SMaSH+ survey. This is particularly relevant for the inner spectroscopic subsystems, where the detection probability depends on a complex interplay of orbital parameters, such as the period ($P_{\mathrm{in}}$), the mass ratio ($q_{\mathrm{in}}$), and the orbital inclination ($i$).

\subsubsection{Spectroscopic and interferometric detection maps}

We estimated the spectroscopic detection probability of unresolved inner binaries by conducting Monte Carlo (MC) simulations following the methodology described in \citet{Sana+2009}. The simulations were designed to reproduce a typical spectroscopic monitoring campaign of Galactic O stars and include a first-order treatment of spectral line blending, which reduces the detectability of systems with large inner mass ratios (see Appendix \ref{app:completeness} for details).

Figure~\ref{fig:completeness2} presents the resulting spectroscopic detection probability map as a function of inner mass ratio and semi-major axis. The spectroscopic sensitivity is near-total for close systems ($\log a_{\mathrm{in}} \lesssim 0.0$) and remains high ($> 80\%$) across the majority of the parameter space occupied by our sample. As expected, the probability drops significantly only towards long periods and low mass ratios, where the RV semi-amplitude $K_1$ falls below our adopted sensitivity threshold. The inclusion of line blending further reduces the sensitivity to nearly equal-mass systems, where the spectral signatures of both components become difficult to disentangle \citep{Bodensteiner+2020,Sana+2025}.

To quantify the sensitivity of our combined PIONIER and NACO/SAM observations, we calculated a 2D detection probability map across the ($\log \tilde{a}_{\mathrm{out}}$, \qout) parameter space following the MC approach described in \citet{Tramper+2026}. For each grid point in separation and mass ratio, we simulated $10^{4}$ orbital configurations with eccentricities drawn from a uniform distribution ($0.0\leq e \leq 0.9$) and random orientations in three-dimensional space. To convert these physical parameters into observable quantities, we applied $H-$band bolometric corrections following the quadratic calibrations for O-type dwarfs provided in Appendix D of \citet{Tramper+2026}. This accounts for the fact that lower-mass companions are  brighter in the infrared than their bolometric luminosities would suggest, thereby extending the sensitivity to lower mass ratios.

The detectability of each synthetic binary was determined using contrast-dependent resolution limits, reflecting the higher angular resolution achievable for low-contrast systems compared to faint companions. We modelled the instrument response using logistic sigmoid functions centred on the survey limits (1–100~mas for PIONIER and 30–100~mas for NACO/SAM), with sensitivity thresholds of $\Delta H\approx4.0$~mag. In Fig. \ref{fig:completeness2}, the detection probability exhibits a distinct increase (towards lower $\Delta$H) in the transition region near $\log a \approx 1.5$. This is the result of the higher sensitivity of NACO/SAM. While PIONIER is constrained by a relatively strict detection limit $\Delta H\approx4$~mag (for a single snapshot), NACO/SAM is theoretically capable of reaching constrasts as deep as $\Delta H\approx 8$. Even with the conservative cuts applied in this study, the broader dynamic range of NACO/SAM ensures a higher detection probability for low-$q$ companions at intermediate separations compared to the PIONIER snapshots. The final probability map represents the fraction of simulated orbits at a mean distance of 1.5~kpc that satisfy the combined detection criteria of both instruments. The detection curve in Fig. \ref{fig:completeness2} serves as the empirical completeness correction for our statistical analysis of the triple star population.

In Fig.~\ref{fig:completeness2} we overlaid the 16 systems from our sample for which both inner periods and individual component masses could be determined (see Tables~\ref{tab:sources_census_spectro} and \ref{tab:sources_census_interfero}). All spectroscopic systems lie well within the high-completeness regime (detection probability $> 80\%$). This confirms that for the typical mass ratios and periods characteristic of massive hierarchical triples, our spectroscopic census is reliable. 

While the absolute sample size is limited to 26 systems (with 16 utilised for detailed mapping), it represents one of the most homogeneous and well-characterised samples of massive hierarchical triples currently available. The high detection probability ($> 80\%$) for the observed parameters suggests that our results are not significantly skewed by non-detections of the inner binary. Consequently, the observed distributions of $q_{\mathrm{in}}$ and $P_{\mathrm{in}}$ can be considered representative of the massive triple population within the SMaSH+ observational volume, providing a robust benchmark for future dynamical and evolutionary models.

\begin{figure}[!ht]
    \centering
    \includegraphics[scale=0.55]{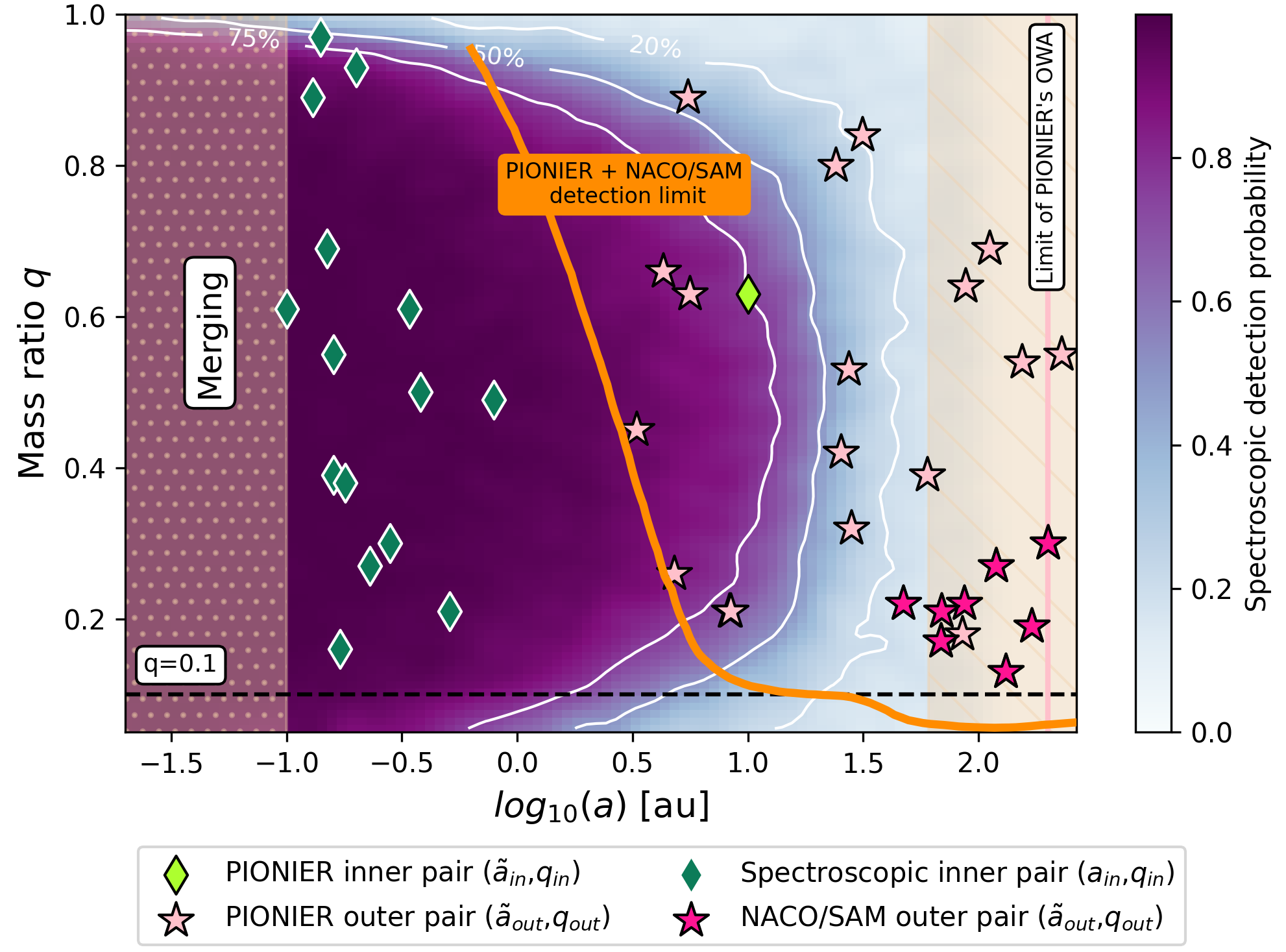}
    \caption{Detection probability map of the inner (resp. outer) spectroscopic (resp. interferometric) subsystems as a function of the inner (resp. outer) semi-major axis (log$_{10}$a, in au) and the mass ratio ($q$). The colour scale indicates the probability of detecting a binary given a $15~\text{km~s}^{-1}$ RV variation threshold. The individual systems from our sample are overplotted:  diamond markers in the case of the inner pair and star markers in the case of outer pairs. For reference, we overlaid the region of PIONIER's OWA (the pink vertical line delimitates the $a_{\rm OWA,80}$ limit), the merging region, and the $q=0.1$ limit.}
    \label{fig:completeness2}
\end{figure}

\subsubsection{Observational biases}
The high angular resolution component of the SMaSH+ survey, primarily conducted with PIONIER and supplemented by NACO/SAM, introduces specific observational biases that define the parameter space box of our sample. 
The primary is the contrast limit, typically $\Delta H < 4$~mag for PIONIER. As shown in the distribution of mass ratios (Fig.~\ref{fig:cdf_q}), our sample is predominantly composed of systems with \qout$ > 0.13$. This is a direct consequence of the dynamic range of the single snapshot approach, lower-mass companions would fall below the detection threshold, particularly around luminous O-type primaries.
Secondly, the angular resolution ($\sim$1.5 to 100~mas) sets the physical boundaries of the outer separations ($a_{\mathrm{proj}}$) shown in our analysis. At the typical distances of our targets ($\sim 1$--$3$~kpc), this corresponds to the $\lesssim 200$~au limit mentioned previously. Within the parameter space defined by angular separations $\rho\approx \tilde{a}_{\mathrm{out}}/d<100$~mas and mass ratios \qout>0.1, our detection map is nearly complete, ensuring that the observed distributions are representative of the intrinsic ones. By design, our study focuses on these relatively tight hierarchical configurations; however, expanding this analysis to include wider separations is possible by incorporating high-contrast imaging data, for example. However, as demonstrated by \citet{Rainot+2022} and \citet{Pauwels+2023}, the majority of the companions at these far out separations consists of low-mass, high-contrast objects that would not meet our $\Delta H < 4$~mag  and \qout$>$0.1 completeness criteria. 
An additional source of incompleteness may affect systems with large outer mass ratios (\qout$>1$). While such companions are readily detected by long-baseline interferometry owing to their low contrast, the identification of an additional spectroscopic subsystem can become more challenging. In nearly equal-luminosity systems, both components contribute significantly to the observed spectrum, leading to increased flux dilution and spectral line blending. This may reduce the measured radial-velocity amplitudes and hinder the detection of unresolved inner companions, particularly when the subsystem is hosted by the secondary component of the outer pair. Consequently, the apparent scarcity of hierarchical systems with (\qout$>1$) in our sample may be partially influenced by observational biases, and should therefore be interpreted with caution.

\subsection{Evolutionary implications and the triple pathway}
\label{sec:evolution}

The architectural properties of our sample, specifically the high frequency of tight inner binaries and relatively massive tertiaries ($q_{\mathrm{out}} > 0.5$), suggest that a significant fraction of massive triples may undergo evolutionary pathways influenced by a third companion. As discussed in the context of dynamical stability (Sect.~\ref{subsect:KL_timescales}), some of our systems currently reside in a regime where general relativistic precession suppresses active ZKL oscillations. However, the prevalence of short-period inner binaries in our sample ($P_{\mathrm{in}} < 10$~d) is consistent with a history of ZKL-driven migration during earlier stages of formation or pre-main-sequence evolution \citep[e.g.][]{Fabrycky+2007,Bataille+2019,Kummer+2023}. 

The long-term fate of these systems is dictated by the presence of the tertiary as the components evolve off the main sequence. In a standard binary scenario, the primary would expand and initiate Roche-lobe overflow (RLOV; e.g. \citealt{Sen+2022}). However, in a hierarchical triple, the increase in stellar radii and associated wind mass loss can trigger dynamical instabilities long before the primary fills its Roche lobe. For systems in our sample with lower separation ratios, such as CPD-47$\degr$2963 or HD~150136, the tertiary may become involved in a common-envelope phase or induce eccentricity pumping in the inner binary as the mass-loss alters the $\tilde{a}_{\mathrm{out}}/a_{\mathrm{in}}$ ratio. This triple-mediated evolution can lead to exotic outcomes, including premature mergers or the formation of stripped-envelope stars with distant companions, which are distinct from the products of isolated binary evolution \citep[e.g.][]{Michaely+2014,Glanz_Perets+2021,Comerford_Izzard_2020,Stegmann+2022,Kummer+2023,Bruenech+2024,Dorozsmai+2024}.

Furthermore, our finding that massive triples often host tertiaries with comparable masses to the inner components has significant implications for the progenitors of gravitational wave sources and the survival of these systems through supernova (SN) events. The likelihood of a system remaining bound depends strongly on the pre-SN orbital configuration and the magnitude of the natal kick. While recent studies propose a wide range of kicks for black holes (BHs) formed from stars with $M > 25\text{--}30~M_{\odot}$ \citep[e.g.][]{Nagarajan+2025}, evidence from low-eccentricity systems like VFTS~243 suggests that some BHs formed via direct collapse may receive only weak kicks \citep{Mahy+2022, Shenar+2022, Willcox+2025}. Such quiet collapses are essential for allowing hierarchical triples to remain bound across multiple evolutionary stages, especially when the tertiary companion is at a significant distance.

The stability of our specific sample during these transitions is underscored by the work of \citet{Kummer+2023}, who show that unbinding of the inner or outer orbit is more probable when the initial periods exceed a few thousand days. The vast majority of the spectroscopic binaries in our sample do not fall into this vulnerable regime, as their inner periods are typically only a few days. One notable exception in our study is HD~101413 (see Table \ref{tab:sources_census_interfero}), the only systems where the inner binary was resolved interferometrically. With a projected separation of $10.0$~au, its period likely exceeds $3000$~days, placing it in a regime that is uniquely susceptible to inner orbital unbinding during a core-collapse event. 

While one can speculate on various formation and evolution scenarios based on these findings, the paths outlined here are intended as potential directions for future study. Ultimately, incorporating the realistic physical and orbital parameter distributions derived in this work into detailed SPH simulations or population synthesis models will be necessary to definitively assess the final fate of individual systems and the population as a whole.

\section{Conclusions}
\label{section:Conclusions}

This study identifies 26 hierarchical O-type massive triple systems within the SMaSH+ survey, providing the first representative set of observed physical and orbital parameter distributions for main-sequence hierarchical massive triples. Our findings provide observational constraints for future population synthesis models of high-mass stars. The key results of this work are summarised as follows:
\begin{enumerate}
    \item The massive triples in our sample are strongly hierarchical, typically consisting of an inner spectroscopic binary and an outer companion detected via interferometry or aperture masking at projected separations of $\approx$ 3–200~au. These systems are found in high stable configurations (in the sense of \citealt{Mardling+2002}), characterised by a median hierarchy ratio ($\beta=\tilde{a}_{out}/a_{in}$) of 127.3 and a minimum observed ratio of 4.34. With nearly 70\% of the systems exhibiting outer-to-inner separations ratio greater than 70, we identified a statistically significant positive correlation between the inner and outer orbital scales. This suggests that the physical processes governing the formation of these systems (e.g. turbulent or disk fragmentation) impose a specific scaling relationship between the resulting fragments. 
    \item We found a broad diversity in the mass of the tertiary companions ($M_\mathrm{3}$). Approximately 38\% of the sample (ten out of 26 systems) host relatively massive tertiaries (\qout$>$0.5). High-mass companions are notably scarce at projected separations below 5 au, an observation that appears to be intrinsic rather than a result of observational bias. We found no significant correlation between the tertiary mass and either the inner-binary mass or the outer separation, indicating that \aout\, and \qout\, can be treated as largely stochastic parameters in population synthesis models.
    \item For several systems with well-constrained orbital solutions, we found that general relativistic precession currently dominates over ZKL oscillations. While ZKL-driven migration likely played a role in the historical evolution and migration of these tight inner binaries, the majority of the five tested systems are currently in a state where secular dynamical oscillations could be suppressed.
    \item We evaluated the observational completeness of our sample and find no substantial detection bias within the parameter space defined by angular separations $\rho<$100~mas (physical separations \aout$\lesssim$200~au) and mass ratios \qout$>$0.1. Our detection map is nearly complete, ensuring that the derived distributions of mass ratios and separations are representative of the intrinsic populations of massive hierarchical triples.
    \item The empirical joint and marginal distributions presented here provide the first observationally grounded initial conditions for the evolution of massive triples. These constraints are essential for understanding the role of tertiary companions in shaping the evolution of massive stars, including their impact on binary interaction rates and the formation of gravitational wave progenitors.

\end{enumerate}

Further monitoring campaigns are essential to refine the current orbital solutions and to enable the characterisation of additional systems, thereby expanding the available sample of massive triples with known orbits. In particular, several systems presented in Appendices \ref{appendix:triples?}, \ref{appendix:list_of_targets}, and \ref{appendix:list_of_quadruples} would benefit significantly from dedicated spectroscopic follow-up to definitively constrain their inner orbital parameters. Future efforts incorporating robust interferometric constraints on outer orbits, alongside realistic distributions for \eout\, and $i_{\text{rel}}$, will be critical for predicting long-term dynamical pathways. Such data are essential to refine our understanding of triple-system evolution and the initial conditions of massive star formation.

\begin{acknowledgements}
This publication is based on observations collected at the European Organisation for Astronomical Research in the Southern Hemisphere.
This project has received funding from the European Research Council under European Union's Horizon 2020 research programme (MULTIPLES, No 772225). EB gratefully acknowledges support by the CRC 1601 (SFB 1601 sub-projects A3) funded by the Deutsche Forschungsgemeinschaft (DFG, German Research Foundation) – grant number 500700252. ST acknowledges support from the Netherlands Research Council NWO (VIDI 203.061 grant). The authors thank the referee for their comments, which helped improve and clarify the manuscript.
This research has made use of the SIMBAD database and VizieR catalogue access tool, operated at CDS, Strasbourg, France. Part of our code made use of Astropy (\url{http://www.astropy.org}) a community-developed core Python package for Astronomy \citep{astropy:2013, astropy:2018}. We used the internet-based NASA Astrophysics Data System (and its new interface SciX) for bibliographic purposes. 
\end{acknowledgements}

\bibliographystyle{aa}
\bibliography{references}

\newpage
\begin{appendix}

\section{Hierarchical triple or massive binary?}
\label{appendix:triples?}

Our study leverages a significant overlap between spectroscopic radial velocity detections and interferometric resolution (see Fig. \ref{fig:completeness2}). While this synergy enables the detection of close massive triples, it also reveals cases where the interferometrically resolved companion and the spectroscopic companion are one and the same. To ensure the hierarchical nature of our sample, we performed a comprehensive literature review of the 73 sources observed with PIONIER and NACO/SAM as part of the SMaSH+ campaign (see Sect. \ref{section:obs}). The primary goal was to distinguish between detected tertiary companions and cases where the interferometric detection simply resolved the known spectroscopic binary (SB) companion.

\subsection{Ambiguous and unconstrained systems}

For systems where an inner spectroscopic orbital solution is available, the distinction is most of the time straightforward. However, for a subset of the SMaSH+ targets, spectroscopic solutions are currently lacking, making it challenging to confirm if the PIONIER detection represents a new hierarchical component or the inner pair itself. This ambiguity mostly concerns long-period binaries where detection in spectroscopy can be challenging due to small RV variations, but where spatial resolution with PIONIER is possible close to its IWA. Consequently, we have excluded these targets from our present study. Representative examples of these excluded targets include HD~152147 \citep[Mahy priv. comm.;][]{Putkuri+2022}, HD~168112, and HD~167659. For other similar cases, we refer to the target notes of \citet{Sana+2014}. Furthermore, systems identified as pair of binaries (2+2 quadruples, e.g. HD~64315) were also discarded to maintain a clean sample of hierarchical triples.

\subsection{Comparison of spectroscopic and interferometric separations}

\begin{figure}[!htb]
\centering
\includegraphics[scale=0.40]{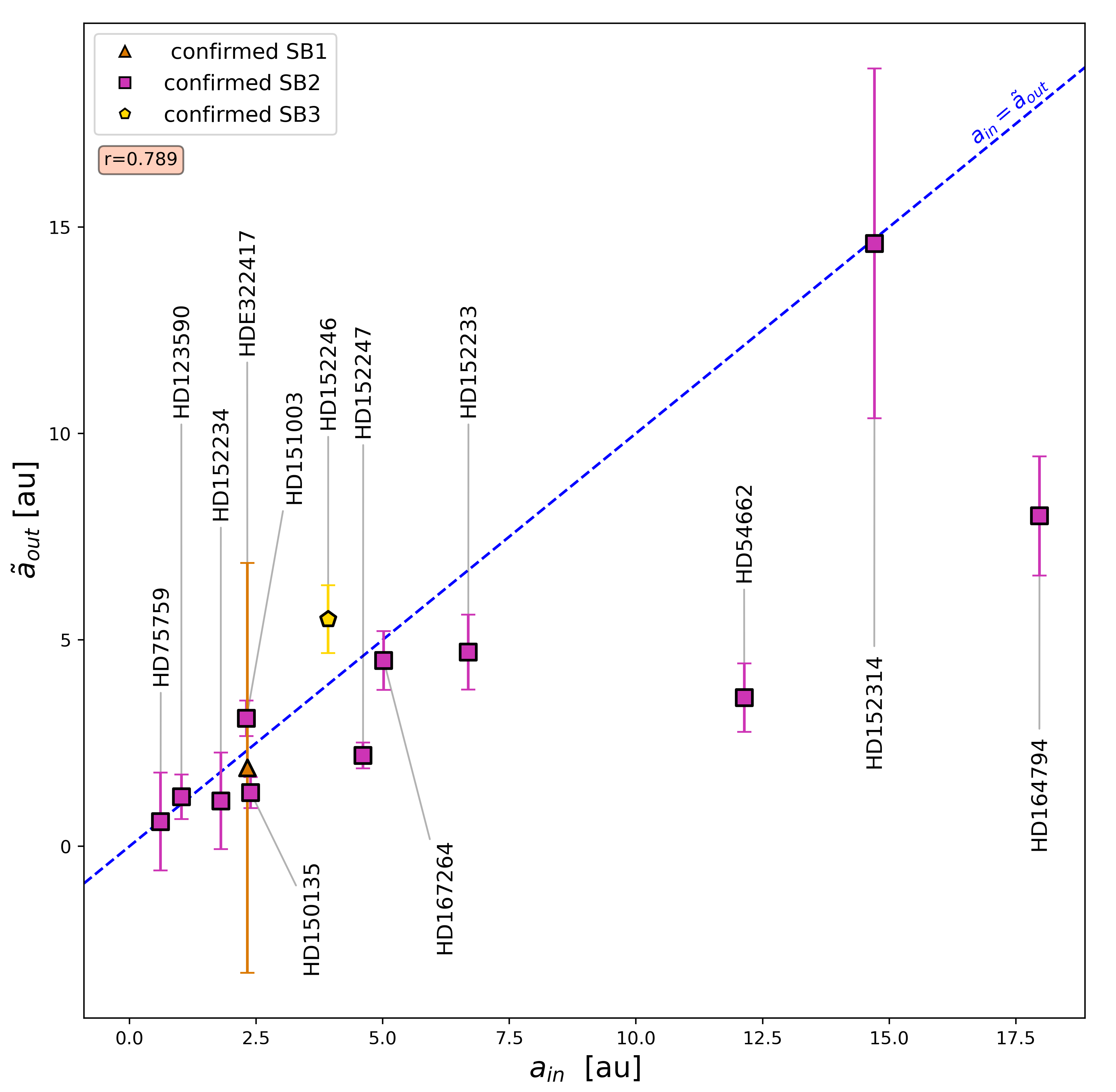}
\caption{Interferometric projected instantaneous separation vs spectroscopic inner separation for a selection of targets with uncertain hierarchical triple configurations. }
\label{fig:triples_or_binaries}
\end{figure}

Figure \ref{fig:triples_or_binaries} displays the spectroscopic separation as a function of the projected interferometric separation for sources with available data. This diagnostic plot allows us to identify systems where $\tilde{a}_{\text{out}}\approx a_{\text{in}}$, suggesting a binary nature rather than a triple.

Two objects, HD~54662 and HD~164794, lie significantly outside the $\tilde{a}_{\text{out}}\approx a_{\text{in}}$ equality curve. Literature insights clarify these cases: \citet{Mossoux+2018} identify HD~54662 as a long-period binary ($P\approx2103.4$~d, $e=0.11$). Similarly, HD~164794 is a confirmed binary with an 8.9-year period \citep{LeBouquin+2017, Fabry+2021}; its high eccentricity ($e=0.648$) likely accounts for the slight discrepancy in its projected separation depending on the orbital phase during the SMaSH+ observations. The high eccentricity of the latter likely accounts for the discrepancy in its projected separation; in highly eccentric orbits, there is a significantly higher probability of detecting companions near apoastron rather than periastron, as the system spends the majority of its orbital period at larger separations.

\subsection{Photometric screening with TESS}
\label{sec:TESS}

To address the possibility of overlooked eclipsing binaries (EB) among the inner systems that are not detected through RV methods, we performed a photometric screening of the 73 targets in our sample observed with PIONIER and/or NACO SAM using the automated variability classification algorithm described in \citet{IJspeert+2024}. This tool is specifically designed to detect periodic signals and eclipses in NASA’s Transiting Exoplanet Survey Satellite (TESS) light curves of massive stars. Of the initial 73 targets, 14 systems turn out to be eclipsing binaries. Nine were already confirmed members of our hierarchical triple sample. HD~150135 and HD~150136 are not resolved by TESS, but the known 2.67-day eclipsing binary in HD~150136 \citep{Mahy+2012,Mahy+2018} allowed us to attribute the photometric signal to that component and subsequently reject HD~150135. We also investigated HD~93160, the only new discovery of this screening, which was previously reported as an SB1 without extensive detail by \citet{Chini+2012}. While the spectroscopic and eclipsing companions may be the same, HD~93160 was ultimately rejected as it appears to be a hierarchical quadruple with more than two companions within 100~mas. TESS finds an EB with a period of 4.3 days. Regarding CPD$59^{\circ}$-2603 ($P_{EB}=1.0749$~d), the system was initially not part of the SMaSH+ sample but was observed as a filler. To maintain consistency with the rest of the triples meeting the SMaSH+ cutoff ($H<7.5)$ and to avoid introducing unnecessary biases into our statistical distributions, we decided not to include it in the final analysis. Full information on this known hierarchical triple system can be found in \citet{Rauw+2001}. A companion is detected with PIONIER at $\rho=(3.43-3.68)\pm(0.15-0.42)$~mas. Finally, TESS detections such as HD~93206 and HD 57061 were already confirmed as higher-order multiples (see Sect. \ref{sec:special_cases}) and excluded. This screening confirms that no low-mass eclipsing pairs were overlooked within the available TESS data, and while the process was necessary to ensure resolved components were not hidden subsystems, TESS did not bring significant additional information to our final sample.

\subsection{Special cases}
\label{sec:special_cases}

HD~93206 (QZ Car): This system exhibits extreme complexity, consisting of at least nine components. It features two bright SBs forming a central quadruple, surrounded by additional members in a trapezium-like configuration \citep{Allison+2011, Sana+2014}.  Similarly, HD~152623 and HD~159176 are of high-multiplicity order within $\sim$1000~au. Such systems represent only 1–2\% of the SMaSH+ data and likely involve different dynamical stability pathways than the hierarchical triples that are the focus of this work. We provide further information on these quadruples in Appendix \ref{appendix:list_of_quadruples}.

While CPD$-47^{\circ}2963$ meets our selection criteria, its nature remains debated. Spectroscopic surveys suggest a 59-day period \citep{Barba+2010, Sota+2014}, whereas interferometric orbit reconstruction suggests a much longer 655-day period \citep{LeBouquin+2017}. The discrepancy (a factor of $\sim$4) suggests the reported SB period may be inexact. We retained this system in our sample but provide further discussion on its parameters in Sect. \ref{section:discussion}.

\section{Inventory of the triple systems and their stellar parameters}
\label{appendix:list_of_targets}

We detail the physical and orbital parameters of the hierarchical triples in our sample. Table \ref{tab:sources_census_spectro} presents the parameters for the majority of the sample (25 systems out of 26), consisting of systems where a tight inner spectroscopic binary is orbited by an outer companion resolved through high angular resolution techniques. Table \ref{tab:sources_census_interfero} provides the same set of parameters for the only resolved triple (1 system out of 26), where both the inner and outer pairs have been spatially resolved via interferometry and aperture masking. It includes target identifiers, spectral types (mostly from the surveys of \citet{Barba+2010} and \citealt{Sota+2014}), magnitudes, inner orbital periods and semi-major axes when available, distances, outer projected separations measured from PIONIER or NACO/SAM detections, and the derived masses and mass ratios for both the inner and outer components. The majority of the masses and distances are adopted from \citet{Tramper+2026}, unless a more detailed study of the system was available, in which case the appropriate reference is provided in the final column. 

We note that 22 of these hierarchical triples have known orbital solutions for the inner pair. For the remaining four systems denoted with an (*) in the table (HD~155806, HD~76556, HD~97253, and HD~76341) the current spectroscopic data are insufficient to reconstruct an inner orbital solution. While their RV data exhibit variability, they do not yet allow   a full orbit reconstruction. Nevertheless, the PIONIER and NACO/SAM detections provide high confidence in the existence of a tertiary companion, although further spectroscopic monitoring is required to characterise the inner systems. Consequently, any analysis or figures involving inner-binary properties, such as the semi-major axis or orbital period, are restricted to the 22 systems with well-constrained orbits.

\begin{table*}[!t]
    \caption{Sample of targets involving an inner spectroscopic binary and their corresponding physical parameters.}
    \scriptsize\addtolength{\tabcolsep}{-3pt}
    \label{tab:sources_census_spectro}
    \begin{tabular}{c c c c c c c c c c c c c c c}
    \hline
    \hline \\ [-1.5ex]
        Target & Spectral Type & $H$ & M$_{H}$ & SB type & P$_{\mathrm{in/out}}$ & $d$ & \ain & $\rho$ & \aout & $\text{M}_{1+2}$ & \textbf{$q_{\text{in}}$} & $\text{M}_{3}$ & $q_{\text{out}}$ & Ref. \\
    [1.5ex]
        & $\text{LC}_{comp}$ & (mag) &  &  & (d) & (pc) & (au) & (mas) & (au) & (\Msun) & $\frac{\text{M}_{2}}{\text{M}_{1}}$ & (\Msun) & $\frac{\text{M}_{3}}{\text{M}_{1+2}}$ & \\

        \hline \\ [-1.5ex]
    
        \multirow{2}{*}{HD~93130} & O6.5III & 8.04 & -4.93 & SB2/SB1 & 23.9 & $2485\pm230$ & 0.52 &  &  & 31.8 & ~ & ~ & ~ & 1,2,G \\ 
        ~ & V & ~ & -2.64 &   &   &  &  & $24.04\pm0.79$ & $59.8\pm5.9$ & ~ & ~ & 12.3 & 0.39 \\ 
        
        \multirow{2}{*}{HD~96670} & O8.5I & 7.07 & -5.62 & SB2/SB1 & 5.3 & $3535\pm51$ & 0.23 &  &  & 31.4 & 0.27 & ~ & ~ & G,3,4 \\
        ~ & IV & ~ & -4.35 &  ~ & ~ &   &  & $29.87\pm0.82$ & $105.6\pm3.3$ & ~ & ~ & 26.5 & 0.84 \\ 
        
        \multirow{2}{*}{HD~101190} & O6IV+O7V & 7.17 & -4.79 & SB2 & 6.05 & $3413\pm477$ & 0.28 &  &  & 48.5 & $>0.3$ & ~ & ~ & 5,6,7 \\
        ~ & V & ~ & -4.17 & ~ & ~ &    &  & $25.73\pm0.6$ & $87.8\pm12.4$ & ~ & ~ & 31.1 & 0.64 \\ 
        
        \multirow{2}{*}{HD~135240} & O7.5IV+B & 5.22 & -3.94 & SB2 & 3.9 & $809.9\pm1.8$ & 0.18 &  &  & 36.36 & 0.55 & ~ & ~ & 8,9,36,G \\
        ~ & V & ~ & -2.2 & ~ & 1602.24 & ~ & ~ & $3.78\pm0.46$ & $3.3\pm0.9$ & ~ & ~ & 16.29 & 0.45 \\ 
        
        \multirow{2}{*}{HD~150136} & O3-3.5V+O5.5-6V+O6.5-7V & 5.09 & -4.79  & SB3 & 2.67 & $1194\pm167$ & 0.17 &  &  & 72.3 & 0.69 & ~ & ~ & 10,11,12,13 \\
        ~ & V & ~ & -3.28 &  ~ & 3144 &  &  & $6.95\pm1.74$ & $8.3\pm2.4$ & ~ & ~ & 15.5 & 0.21 \\ 
        
        \multirow{2}{*}{HD~158186} & O9.5V & 6.88 & -3.25 & SB2/SB1 & 2.51 & $1045\pm334$ &  0.10 & & & 17.4 & ~ & ~ & ~ & 2 \\ 
        ~ & V & ~ & -1.12 & ~ & ~ & ~ &  &  $26.9\pm0.38$ & $28.1\pm9.0$ & ~ & ~ & 5.5 & 0.32 \\ 
        
        \multirow{2}{*}{HD~159176} & O7V+O7V & 5.51 & -3.94 & SB2 & 3.37 & $1099\pm261$ &  0.14 & & & 26.7 & 0.97 &  ~ & ~ &  14 \\
        ~ & V & ~ & -1.55 & ~ & ~ & ~ & & $64.13\pm1.76$ & $4.8\pm1.1$ & ~ & ~ & 6.9 & 0.26 \\ 
        
        \multirow{2}{*}{HD~167971} & O8Ia+(O6.5V+O6.5V) & 5.32 & -5.62 & SB3 & 3.2 & $1610\pm5.7$ & 0.19 & &  & 62.3 & 0.93 & ~ &  ~ & 13,M,15,16 \\
        ~ & IV & ~ & -5.53 & ~ & 7806 & ~ & & $17.02\pm0.38$ & $27.4\pm0.6$ & ~ & ~ & 33.4 & 0.53 \\ 
        
        \multirow{2}{*}{CPD$-47\degr2963$} & O5I & 6.06 & -5.63 &  SB(1,2,3?) & 59 & $2109\pm41$ & 1.3 & & & 50.3 & ~ & ~ & ~ & 13,O,G,17 \\
        ~ & V & ~ & -4.21 & ~ & 655 & ~ & & $2.64\pm0.13$ & $5.6\pm0.3$ & ~ & ~ & 31.9 & 0.63 \\ 
        
        \multirow{2}{*}{HD~101131} & O5.5V+O8V & 7 & -4.36 & SB2 & 9.7 & $2528\pm476$ & 0.34 & & & 35.5 & 0.61 & ~ & ~ & 6,18 \\
        ~ & V & ~ & -3.41 & ~ & ~ & ~ & & $45.45\pm1.12$ & $154.4\pm29.3$ & ~ & ~ & 19.2 & 0.54 \\ 
        
        \multirow{2}{*}{HD~152723} & O6.5III & 6.74 & -4.93 & SB2 & 18.9 & $2192\pm203$ & 0.51 & & & 31.8 & 0.21 & ~ & ~ &  O,G,20 \\ 
        ~ & V & ~ & -3.24 & ~ & ~ & ~ & & $80.75\pm1.0$ & $229\pm21.4$ & ~ & ~ & 17.4 & 0.55 \\ 
        
        \multirow{2}{*}{HD~155806$^{(*)}$} & O7.5V & 5.69 & -3.8 & SB2 & ~ & $969\pm247$ & ~ & ~ &  & 24.4 & ~ & ~ & ~ & 17,21 \\ 
        ~ & V & ~ & -3.43 & ~ & ~ & ~ & & $24.87\pm0.68$ & $24.1\pm6.1$ & ~ & ~ & 19.4 & 0.8 \\ 

         \multirow{2}{*}{HD~164816} & O9.5V+B0V & 7.05 & -3.25  & SB2 & 3.8 & $1488\pm476$ & 0.13 & & & 17.4 & 0.89 & ~ & ~ & O,G,22,23,24 \\
        ~ & V & ~ & 0.05  & ~ & ~ & ~ & & $56.93\pm2.06$ & $85.2\pm27.4$ & ~ & ~ & 3.1 & 0.18 \\

        \multirow{2}{*}{HD~167263} & O9.5III & 5.91 & -4.52 & SB2 & 14.76 & $1324\pm184$ &  1.03 & & & 20.4 & ~ & ~ & ~ & 3,G,17,21,25 \\
        ~ & IV & ~ & -3.52  & ~ & ~ & ~ & & $79.3\pm0.43$ & $111.6\pm15.5$ & ~ & ~ & 14.1 & 0.69 \\ 

        \multirow{2}{*}{HD~76556$^{(*)}$} & O6IV & 7.14 & -4.63 &  SB2 & ~ & $1950\pm197$ & ~ & ~ &  & 33.3 &~ & ~ & ~ & G\\ 
        ~ & V & ~ & -1.56 & ~ & ~ & ~ & & $4.31\pm0.62$ & $8.4\pm1.5$ & ~ & ~ & 7 & 0.21 \\

        \multirow{2}{*}{HD~97253$^{(*)}$} & O5III & 6.712 & -5.14 & SB1 & ~ & $2260\pm159$ & ~ & ~ & & 40.2 & ~ & ~ & ~ & 27,53,G\\ 
        ~ & V & ~ & -3.19 & ~ & ~ & ~ & & $11.24\pm0.55$ & $25.4\pm2.2$ & ~ & ~ & 16.8 & 0.42 \\

       \multirow{2}{*}{ HD~115455$^{(N)}$} & O8III & 7.4 & -4.73 & SB2 & 15.1 & $2479\pm287$ & 0.38 & ~ & & 25.4 & 0.50 & ~ & ~ & 25, OWN\\ 
        ~ & V & ~ & -1.51 & ~ & ~ & ~ & & $48.0\pm9.2$ & $119.3\pm8.7$ & ~ & ~ & 6.8 & 0.27 \\

        \multirow{2}{*}{ HD~152219$^{(N)}$} & O9III+B1-2VIII & 7.2 & -4.59 & SB2 & 4.2 & $2384\pm313$ & 0.16 & ~ & & 25.9 & 0.39 &  ~ & ~ & 29\\ 
        ~ & V & ~ & -1.75 & ~ & ~ & ~ & & $83.6\pm9.2$ & $199.3\pm34.1$ & ~ & ~ & 7.7 & 0.30 \\

        \multirow{2}{*}{ HD~152246} & O9IV+O9 & 6.8 & -4.01 & SB3 & 6.0 & $1741\pm255$ & 0.23 & ~ & & 22.78 & ~ & ~ & ~ & 30 \\ 
        ~ & IV & ~ & -3.7 & ~ & 470 & ~ & & $3.34\pm0.16$ & $5.5\pm0.8$ & ~ & ~ & 20.35 & 0.89 \\

         \multirow{2}{*}{ CPD$-41\degr7733^{(N)}$} & O8.5+B3 & 7.46 & -3.52 & SB2 & 5.7 & $1604\pm461$ & 0.18 & & & 20.5 & 0.38 & ~ & ~ &  31 \\
        ~ & V & ~ & -0.2 & ~ & 655 & ~ & & $42.7\pm21.36$ & $68.5\pm39.5$ & ~ & ~ & 3.5 & 0.17 \\

        \multirow{2}{*}{ HD~165246$^{(N)}$} & O8V+B7V & 7.3 & -3.66 & SB2 & 4.59 & $1547\pm419$ & 0.17 & ~ & & 27.5 & 0.16 & ~ & ~ & 32, 8 \\ 
        ~ & V & ~ & -1.3 & ~ &  & ~ & & $30.5\pm16.07$ & $47.2\pm28.0$ & ~ & ~ & 6.1 & 0.22 \\

        \multirow{2}{*}{ HD~168075$^{(N)}$} & O6.5V+B0-1V & 7.4 & -4.08 &  SB2 & 43.6 & $1963\pm435$ & 0.78 & ~ & & 29.3 & 0.49 & ~ & ~ & 29, OWN\\ 
        ~ & V & ~ & -0.38 & ~ &  & ~ & & $44.14\pm27.02$ & $86.7\pm56.4$ & ~ & ~ & 6.1 & 0.22 \\

        \multirow{2}{*}{ HD~47129$^{(N)}$} & O8III+07.5III & 5.8 & -5.3 &  SB2 & 14.4 & $1898\pm195$ & 0.37 & ~ & & 28 & 1.04 & ~ & ~ & 33\\ 
        ~ & V & ~ & -0.38 & ~ &  & ~ & & $36.44\pm18.64$ & $69.2\pm36.1$ & ~ & ~ & 5.8 & 0.21 \\

        \multirow{2}{*}{HD~76341$^{(*,N)}$} & O9.5IV & 6.4 & -3.9 &  SB2 & & $1008\pm155$ & ~ & ~ & & 18.7 & ~ & ~ & ~ & GOSSS\\ 
        ~ & V & ~ & -0.18 &  ~ &  & ~ & & $168.9\pm8.46$ & $170.3\pm27.5$ & ~ & ~ & 3.5 & 0.19 \\

        \multirow{2}{*}{HD~164740} & O9.5V+B07V & 7.4 & -3.8 &  SB3 & 1.54 & $1234\pm16$ & 0.1 & ~ & & 33.0 & 0.61 & ~ & ~ & 34,35 \\ 
        ~ & O7.5V & ~ & ~ & ~ & 496.82 & ~ & & $3.49\pm1.13$ & $4.30\pm1.26$ & ~ & ~ & 22.3 & 0.66 \\
        
        \hline

    \end{tabular}

\tablefoot{We provide for each source its spectral type and the luminosity class of the detected companion. Column 3 to 5 respectively gives the H-band magnitude ($H$), the derived absolute magnitude (M$_{H}$), and the V-band magnitude ($V$). SB Type stands for the spectroscopic binary type. We adopted the terminology consistent with that used in the graph presented in the paper; the more precise SB type (including the eclipsing nature) can be found in the references. For the inner binaries with orbital solutions, we provide the period in the next column. The distance is given in parsecs and the angular separation (Ang. Sep.) in milliarcseconds. Proj. Sep. stands for projected separation and is provided in astronomical units (au). Columns 11-13 respectively provide the mass of the inner binary ($\text{M}_{1+2}$), the mass of the interferometric companion ($\text{M}_{3}$), and the mass ratio ($\text{q}_{out}$). Angular separations are  from \citet{Sana+2014}, while the absolute magnitudes and the derived masses are from \citet{Tramper+2026}. Any other reference relevant to fill this table is provided in the last column and is listed below. Targets marked with an exponent (N) indicate cases where the tertiary companion was detected through NACO/SAM according to the criteria in Sect. \ref{section:obs}, whereas all other companions were detected with PIONIER. Targets marked with an asterisk ($*$) denote sources where spectroscopic data are currently insufficient and only RV variability has been detected; for these, PIONIER/NACO detections provide high confidence in the presence of a tertiary, though further spectroscopic investigation is required.}

\tablebib{(1)~\citet{Otero+2006}, (2)~\citet{Avvakumova+2013}, (3)~\citet{Stickland+2001}, (4)~\citet{Gomez+2021}, (5)~\citet{Sana+2011}, (6)~\citet{Pozo+2019}, (7)~\citet{SanaJamesGosset+2011}, (8)~\citet{Mayer+2014},(9)~\citet{Penny+2001}, (10)~\citet{Sana+2013_HD15}, (11)~\citet{Niemela+2005}, (12)~\citet{Mahy+2018}, (13)~\citet{LeBouquin+2017}, (14)~\citet{Linder+2007}, (15)~\citet{Leitherer+1987}, (16)~\citet{Ibanoglu+2013}, (17)~\citet{Martins+2018}, (18)~\citet{Gies+2002}, (19)~\citet{SanaJamesGosset+2011}, (20)~\citet{Mahy+2022}, (21)~\citep{Negueruela+2004}, (22)~\citet{Chini+2012}, (23)~\citet{Mayer+2017}, (24)~\citet{Trepl+2012}, (25)~\citet{Mayer_OB+2014}, (26)~\citet{Stickland+1998}, (27)~\citet{Chini+2012}, (28)~\citet{Walborn+1973}, (29)~\citet{Sana+2009}, (30)~\citet{Nasseri+2014}, (31)~\citet{Sana+2007}, (32)~\citet{Johnston+2021}, (33)~\citet{Linder+2008}, (34)~\citet{Sanchez-Bermudez+2022}, (35)~\citet{Campillay+2019}, (36)~\citet{Svrckova+2026} \\ O: OWN~\citet{Barba+2010,Barba+2017},
G: GOSSS~\citet{Sota+2014,MaizAp+2016}, M: MONOS~\citet{MaizApe+2019,MONOS2}}

\end{table*}

\begin{table*}[!t]
    \caption{Physical parameters for HD~101413, the fully resolved triple in our sample. }
    \scriptsize\addtolength{\tabcolsep}{-3pt}
    \label{tab:sources_census_interfero}
    \begin{tabular}{c c c c c c c c c c c c c c c }
    \hline
    \hline \\ [-1.5ex]
        Target & Spectral Type & $H$ & M$_{H}$ & $V$ & $d$  & $\rho$ & $\tilde{a}_{\text{in}}$ & \aout & $\text{M}_{1}$ & $\text{M}_{2}$ & $\text{M}_{3}$ & $\text{q}_{\text{in}}$ & $\text{q}_{\text{out}}$  & Ref. \\
    [1.5ex]
        & $\text{LC}_{\mathrm{comp}}$ & (mag) &  (mag) &  & (pc) & (mas) & (au) & (au) & (\Msun) & (\Msun) & (\Msun) & $\frac{\text{M}_{2}}{\text{M}_{1}}$ & $\frac{\text{M}_{3}}{\text{M}_{1+2}}$ & \\

        \hline \\ [-1.5ex]

     \multirow{3}{*}{HD~101413} & O8V & 8.13 & $-$3.66 & 9.621& $2452\pm665$ & ~ & ~ & & 26.8 & ~ & ~ & ~ & ~  & \multirow{3}{*}{\citet{Putkuri+2026,SanaJamesGosset+2011}}\\ 
    ~ & B3V & ~ & $-$2.21 & ~ & ~ &  $4.1\pm0.8$ & $10.0\pm1.9$ & ~ & ~ &  14.7 & ~ & 0.63 & & \\
    ~ & V & ~ & $-$1.07 & ~ & ~ & $53.62\pm7.58$ & ~ & $131.5\pm40.2$ & ~ & ~ & 5.4 & ~ & 0.13 &  \\
    \hline

    \end{tabular}

\tablefoot{The parameters are identical to those in Table \ref{tab:sources_census_spectro}, with the addition of $\tilde{a}_{\text{in}}$, the inner projected separation derived from PIONIER data. The primary of HD~101413 was detected via PIONIER, while the secondary was detected through NACO/SAM.}
\end{table*}

\section{Inventory of the quadruple systems and their stellar parameters}
\label{appendix:list_of_quadruples}

Four quadruple systems are currently known within the SMaSH+ sample. Three of them, HD~152623, HD~57061, and HD~93206, were already identified in previous studies, while the fourth system, HD~93160, was uncovered through our TESS eclipsing-binary check of the present study. Their main properties are summarised in Table~\ref{tab:quadruples}.

Although 2+2 quadruples can, in principle, be interpreted as two coupled hierarchical triples, and 3+1 configurations may experience additional dynamical evolution of the intermediate orbit through ZKL mechanisms, we restricted the statistical analysis in this study to hierarchical triple systems. A comprehensive treatment of quadruples would require modelling their full architectures and mutual interactions, which is beyond the scope of this work, particularly given that orbital parameters of the individual subsystems are often incomplete or poorly constrained.

For representation, each quadruple system was reduced to an effective triple by identifying the most relevant inner binary and treating the remaining resolved component(s) as a hierarchical, outer tertiary companion. In most cases, the resolved companions were therefore assigned as outer tertiaries orbiting the inner eclipsing or spectroscopic binary. For HD~152623, only one hierarchical configuration was retained: although two resolved companions are present within 1000~au, the second companion lies at a projected separation of $\sim$445~au, and is therefore outside the separation range probed in this study with PIONIER and NACO (i.e. within $\sim$200~au). This component was thus not included in the plots. For HD~93206, which is a known 2+2 system, we adopted a simplified representation in which the resolved pair Ac1–Ac2 is treated as a single effective tertiary component orbiting the inner Aa1–Aa2 binary. We emphasise that this construction is necessarily artificial and does not reflect the full dynamical complexity of the system.

These systems are shown with black symbols in the figures describing the observed sample, but they are excluded from all derived distributions and statistical analyses. They are included solely as a reference for future studies of higher-order multiplicity in massive stars. We further note that some of these objects, in particular HD~93206 and HD~57061, are part of even higher-order hierarchical configurations when companions at separations larger than 1000~au are taken into account, highlighting the strong dynamical complexity of these systems \citep[see e.g.][]{Sana+2014, Rainot+2020}.

\begin{table*}[!t]
\centering
\small
\caption{Known quadruple systems in the SMaSH+ sample (within 1000~au).}
\label{tab:quadruples}
\begin{tabular}{p{1.5cm} p{1.5cm} p{3.cm} p{3.5cm} p{5.2cm} p{1.2cm} }
\hline
\hline \\[-1.5ex]

Target & Distance (pc) & Spectroscopic \newline companion(s) & Resolved companion(s) \newline ($<1000$~au) &  Comments & Ref. \\ 
\hline \\[-2.0ex]
\multirow{4}{*}{HD~93160} & \multirow{4}{*}{$2543\pm255$} & 
\multirow{2}{3.5cm}{$P_{\rm EB}=4.3$~d}  &
$\tilde{a}_{\rm out,3}=16.4\pm1.7$~au & \multirow{4}{5.2cm}{EB from TESS screening + 2 resolved companions with PIONIER} &
\multirow{4}{1.5cm}{1,2} \\
 & & \multirow{2}{3.5cm}{$M_{\rm EB}=29.5$~\Msun }& $\tilde{a}_{\rm out,4}=79.1\pm8.0$~au &  & \\
 & & & $M_{\rm 3}=19.1$~\Msun & & \\
 & & & $M_{\rm 4}=5.5$~\Msun & & \\[2pt]

\hline \\[-2.0ex]

 \multirow{4}{*}{HD~152623} & \multirow{4}{*}{$1773\pm422$} & 
\multirow{2}{3.5cm}{$P_{\rm SB1}=3.9$~d}  &
$\tilde{a}_{\rm out,3}=50.0\pm0.6$~au & \multirow{4}{5.2cm}{SB1 + 1 resolved companions with CHARA/NACO + 1 resolved companion with PIONIER} &
\multirow{4}{1.5cm}{1,2,3,4,5} \\
 & & \multirow{2}{3.5cm}{$M_{\rm SB1}=26.7$~\Msun }& $\tilde{a}_{\rm out,4}=445\pm10$~au &  & \\
 & &  & $M_{\rm 3}=16.1$~\Msun & & \\
 & &  & $M_{\rm 4}=26.5$~\Msun & & \\[3pt]

 \hline \\[-2.0ex]

 \multirow{3}{*}{HD~57061} & \multirow{3}{*}{$1234\pm146$} &
$P_{\rm EB}=1.28$~d  &
$\tilde{a}_{\rm out,4}=114\pm2$~au & \multirow{4}{5.2cm}{3+1 quadruple: EB confirmed by TESS + SB1 + 1 resolved companion with PIONIER/NACO} &
\multirow{4}{1.5cm}{1,2,6,7,8,9} \\
  & & $P_{\rm SB1}=152$~d & $M_{\rm 4}=21.0$~\Msun  &  & \\
 $\equiv \tau$ CMa & & $M_{\rm EB+SB1}=24.2$~\Msun & & & \\[8pt]

 \hline \\[-2.0ex]

\multirow{5}{*}{HD~93206} & \multirow{5}{*}{$1841\pm29$} &
$P_{\rm Aa12}=20.7$~d  &
$\tilde{a}_{\rm Aa12-Ac12}= 48\pm1 $~au & \multirow{4}{5.2cm}{2+2 quadruple: 2 spectroscopic binaries (Aa1-Aa2 + Ac1-Ac2). Ac12 is an EB from TESS screening. PIONIER resolves the two binaries Aa12-Ac12 } &
\multirow{6}{1.5cm}{1,2,10,11} \\
 & & $P_{\rm Ac12}=5.99$~d &  &  & \\
 & & $M_{\rm Aa1}=43\pm6$~\Msun  & & & \\
  $\equiv$ QZ Car &  & $M_{\rm Aa2}=19\pm3$~\Msun & & & \\
 & & $M_{\rm Ac1}=25$~\Msun & & & \\
 & & $M_{\rm Ac2}=32.2$~\Msun &  &  & \\[2pt]

\hline
\end{tabular}

\tablefoot{EB and SB denote eclipsing binaries and spectroscopic binaries, respectively. Orbital periods are given in days. The separations of the resolved companions correspond to instantaneous projected separations, $\tilde{a}_{\rm out}$, expressed in au. Masses are given in units of solar masses (\Msun) and distances in pc. The distances are   from \citet{Tramper+2026}. For HD~93206, we refer to \citet{Mayer+2022} and \citet{Sana+2014} for a detailed description of the system architecture and its higher-order multiplicity.}

\tablebib{(1)~\citet{Sana+2014}, (2)~\citet{Tramper+2026}, (3)~\citet{Fullerton+1990}, (4)~\citet{Mason+1998}, (5)~\citet{Mason+2009}, (6)~\citet{vanLeeuwen+1997},(7)~\citet{Stickland+1998},(8)~\citet{Tokovinin+2010},(9)~\citet{Burssens+2020},(10)~\citet{Walker+2017},(11)~\citet{Mayer+2022}}

\end{table*}

\section{Dynamical stability criteria}
\label{app:dyn_criteria}

To guarantee the long-term dynamical stability of a triple system, a specific hierarchical arrangement is needed. The assessment of stability is mainly a result of the mass of the objects and their orbital parameters (e.g. separations, eccentricity, relative inclination). However, setting the boundary between a stable and unstable system presents some complexities. During their evolution, triple systems may remain stable for extended periods before transitioning to a state of dynamical instability, ultimately breaking down into a lower-order configuration \citep{VandenBerk+2007}. Several stability criteria exist, but we limited ourselves to the theoretical equations of \citet{Mardling+2002} and the empirical models of \citet{Tokovinin+2004}. The hierarchical dynamical stability is ensured if the separation ratio is greater than a certain critical value $\beta$ defined as follows by \citet{Mardling+2002}:
\begin{equation}
    \frac{\tilde{a}_{\text{out}}}{a_{\text{in}}} > \beta = \left.\frac{\tilde{a}_{\text{out}}}{a_{\text{in}}}\right|_{\text{crit}}=\frac{2.8}{1-e_{\text{out}}}\left(1-\frac{0.3\,i_{\text{rel}}}{\pi} \right) \left[\frac{(1+q_{\text{out}})(1+e_{\text{out}})}{\sqrt{1-e_{\text{out}}}} \right]^{\frac{2}{5}}
    \label{eq:mardling_criteria}
.\end{equation}

For a given value of \eout\, and \qout, a triple configuration with $\frac{\tilde{a}_{\text{out}}}{a_{\text{in}}} < \beta$  indicates that the system encounters dynamical instability and is likely to undergo the ejection of one of its components. In the case of an equal-mass system with a coplanar circular orbit ($q_{\text{out}}=0.5$, \,\, $i_{\text{rel}}=0$ and $e_{\text{out}}=0$, hereafter: Case 1), Eq. \ref{eq:mardling_criteria} simplifies into $\frac{\tilde{a}_{\text{out}}}{a_{\text{in}}} = 3.3$, meaning that the outer companion should lie at a distance greater than 3.3 times the semi-major axis of the inner binary to be in a stable configuration. 

By studying over a hundred multiples of solar-type primaries for which reliable orbital solutions exist, \citet{Tokovinin+2004} measured an empirical stability limit of

\begin{equation}
    \left.\frac{\mathrm{P}_{\text{out}}}{\mathrm{P}_{\text{in}}}\right|_{\text{crit}}=\frac{5}{(1-e_{\text{out}})^{3}}
    \label{eq:Toko}
,\end{equation}

\noindent which can be straightforwardly translated into a ratio of separations using Kepler's third  law (see Eq. \ref{eq:kepler}):

\begin{equation}
    \left.\frac{\tilde{a}_{\text{out}}}{a_{\text{in}}}\right|_{\text{crit}}=\frac{5^{\frac{2}{3}}}{(1-e_{\text{out}})^{2}}\,q_{\text{out}}^{\frac{1}{3}}
    \label{eq:quasi_Tokovinin}
.\end{equation}

Case 1 then leads to $\frac{\tilde{a}_{\text{out}}}{a_{\text{in}}} > 2.3$. 
Both stability criteria are of the same order of magnitude. Still, the empirical model of \citet{Tokovinin+2004} appears slightly less restrictive than the theoretical models. 

\section{Joint distribution of total mass and outer separation}
\label{app:joint_distrib}

\begin{figure}[!ht]
\centering
\includegraphics[scale=0.45]{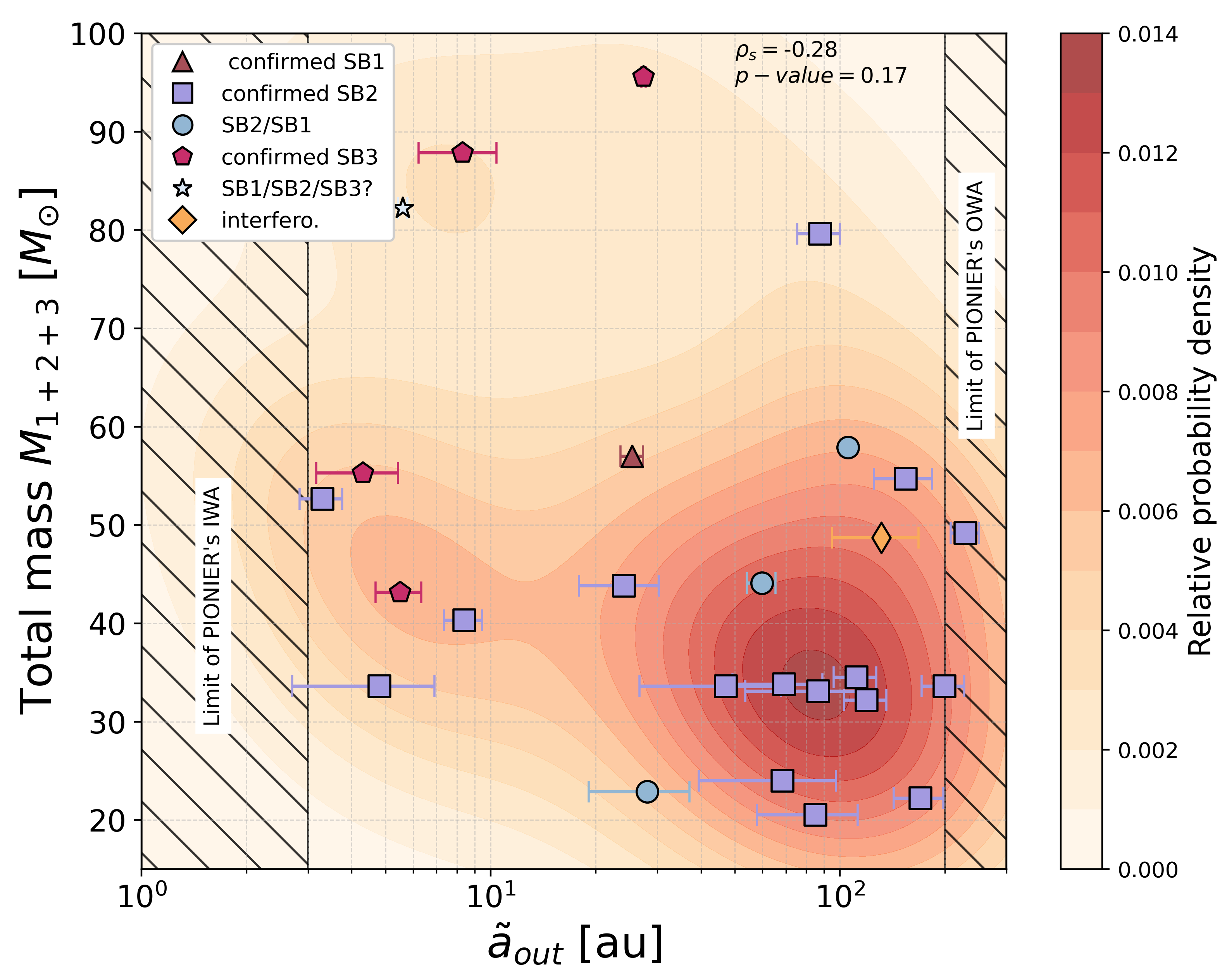}
\caption{Joint PDF of total system mass $M_{1+2+3}$ and mass ratio \qout. The contours show the relative probability density and the different symbols indicate the different types of inner binaries. For reference, we overlay the approximate PIONIER IWA and OWA limits.}
\label{fig:joint_mtot_a}
\end{figure}

In Fig. \ref{fig:joint_mtot_a} we present $f(\tilde{a}_{out},M_{1+2+3})$ the joint probability density function of the total mass of the systems versus the projected outer separation. This distribution explores whether the total mass available in the inner binary influences the eventual distance of the third body. As show in Sect. \ref{sect:joint_distributions}, here again the axis-aligned morphology of the KDE contours reinforces our finding that \aout\, is a stochastic parameter. In turbulent fragmentation models, \citep[e.g.][]{Offner+2014}, the distance between the fragments is determined b y the local Jean's length and turbulent Mach numbers rather than the total mass eventually accumulated by the stars. The lack of correlation ($\rho_{s}=-0.28, p=0.17$) allows population synthesis models to sample $M_{\mathrm{tot}}$ ($=M_{1+2+3}$) and \aout\, independently, and following the marginal distributions derived in Sect. \ref{sect:distribution_q_and_m}. 

\section{Completeness maps and simulations}
\label{app:completeness}

The spectroscopic detection probability was estimated through Monte Carlo simulations designed to provide an approximate characterisation of the sensitivity of our sample to unresolved inner binaries. Because the spectroscopic information was compiled from heterogeneous literature sources, the exact observational properties (number of epochs, time baseline, and radial-velocity precision) vary from target to target and are not uniformly available. Consequently, the simulations do not attempt to reproduce the observational history of each individual system. Instead, they provide a representative estimate of the average spectroscopic detection efficiency for massive binaries occupying different regions of the $(a,q)$ parameter space.

We generated a population of $10^{5}$ synthetic binaries with primary masses randomly distributed (flat distribution) between 15 and 60~\Msun. Mass ratios were drawn from a uniform distribution between 0.1 and 1.0 \citep{Sana+2012}, eccentricities were sampled uniformly between 0 and 0.9, and orbital inclinations were assumed to be isotropically distributed. Semi-major axes were drawn from a uniform logarithmic distribution spanning ($-1.8 \leq \log_{10}(a/{\rm au}) \leq 2.0$). Orbital periods were computed using Kepler's third law.

To mimic a typical long-term spectroscopic monitoring campaign of Galactic O stars, each simulated system was assigned a random number of observing epochs between 6 and 15 distributed over a 5-year baseline. Radial velocities were computed at each epoch by solving the Keplerian orbit, and Gaussian noise with a standard deviation of $3\,{\rm km\,s^{-1}}$ was added to represent measurement uncertainties. A binary was considered spectroscopically detected when the maximum radial-velocity variation measured during the simulated campaign exceeded $20\,{\rm km\,s^{-1}}$, a threshold comparable to those commonly adopted in multiplicity studies of massive stars \citep[see e.g.][]{Bodensteiner+2020,Sana+2025}.

A key limitation of many simplified completeness calculations is the neglect of spectral line blending. In massive binaries, the spectral lines of both components can overlap when their velocity separation becomes comparable to the intrinsic line width of the stellar absorption features. This effect is particularly important for systems with nearly equal masses, for rapidly rotating stars, and for binaries whose orbital velocities are only marginally larger than the line widths. In such cases, the measured radial velocity corresponds to the centroid of the blended profile rather than to the true velocity of the primary star, leading to an attenuation of the observed radial-velocity amplitude and therefore a lower detection probability.

To account for this effect, we adopted a simplified blending prescription in which the luminosity ratio scales as $L_{2}/L_{1} \propto q^{3}$, appropriate for massive stars. At each epoch, the velocity separation $\Delta v = |RV_{1}-RV_{2}|$ was compared to a characteristic line width of $100\, { \rm km \, s^{-1}}$. The observed radial velocity was then interpolated between the true primary-star velocity and the flux-weighted centroid of the blended profile using a blending weight that increases as the velocity separation decreases. This prescription captures the expected reduction in spectroscopic sensitivity for systems with large mass ratios and partially blended spectral lines.

In these systems, the photospheric flux of the outer companion can further dilute the radial-velocity signal of the unresolved subsystem, making spectroscopic detection substantially more difficult than predicted by purely Keplerian simulations. The effect is strongest for outer binaries with $q_{\rm out}\approx1$, where both stars contribute comparable fluxes to the observed spectrum. Consequently, the observed scarcity of hierarchical systems with large outer mass ratios may be influenced, at least in part, by observational selection effects rather than reflecting an intrinsic absence of such systems.

The resulting completeness map (see Fig. \ref{fig:completeness2}) should therefore be interpreted as a representative estimate of the spectroscopic sensitivity of the survey. While it incorporates the first-order impact of observational cadence and line blending, it does not reproduce the exact observing strategy of each target and should not be regarded as a full bias-correction analysis comparable to dedicated survey simulations.

\end{appendix}
\end{document}